\documentclass[conference,letterpaper,twoside,final,10pt]{IEEEtran}

\usepackage{cite}
\usepackage{amssymb}
\usepackage{amsmath}
\usepackage{color}
\usepackage{multirow}
\usepackage{eurosym}
\usepackage{graphicx}
\usepackage[caption=false]{subfig} 
\usepackage{threeparttable}
\usepackage[siunitx]{circuitikz}
\usepackage{gensymb}
\usepackage{flushend}
\usepackage[draft=true]{hyperref}

\newtheorem{observation}{Observation}

\newtheorem{corollary}{Corollary}

\IEEEoverridecommandlockouts

\begin{document}

\title{Multi-hop Backscatter Tag-to-Tag Networks}


\author{\IEEEauthorblockN{Amjad Yousef Majid, Michel Jansen, Guillermo Ortas Delgado, Kas{\i}m Sinan Y{\i}ld{\i}r{\i}m, and Przemys{\l}aw Pawe{\l}czak}\IEEEauthorblockA{Delft University of Technology, The Netherlands
\\\{m.jansen-2,\,g.ortasdelgado\}@student.tudelft.nl,\,\{a.y.majid,\,k.s.yildirim,\,p.pawelczak\}@tudelft.nl
}}

\maketitle

\begin{abstract}
We characterize the performance of a backscatter tag-to-tag (T2T) multi-hop network. For this, we developed a discrete component-based backscatter T2T transceiver and a communication protocol suite. The protocol composed of a novel (i) flooding-based link control tailored towards backscatter transmission, and (ii) low-power listening MAC. The MAC design is based on the new insight that backscatter reception is \emph{more energy costly} than transmission. Our experiments show that multi-hopping extends the coverage of backscatter networks by enabling longer backward T2T links (tag far from the exciter sending to the tag close to the exciter). Four hops, for example, extend the communication range by a \emph{factor of two}. Furthermore, we show that dead spots in multi-hop T2T networks are far less significant than those in the single-hop T2T networks. 
\end{abstract}


\section{Introduction}
\label{sec:introduction}
	
\bstctlcite{BSTcontrol}Backscatter enables ultra low-power communication by eliminating energy-hungry hardware components~\cite{talla_arxiv_2017}. Backscatter radios have small footprints, lower cost and notably several orders of magnitude less energy requirements than their active counterparts~\cite[Sec. 1]{zhang_sigcomm_2016}. Disparity is striking, e.g. a backscatter-based WiFi module is $\approx$10$^\text{4}$ times more energy efficient than an active one~\cite[Fig. 3]{talla_arxiv_2017}. Despite significant energy savings, compared to classical wireless sensor networks, backscatter networks, such as networks of RFID tags, are controlled by a central carrier generator (a reader)~\cite{alevizos_arxiv_2017,griffin_apm_2009} allowing \emph{only} one-to-many (single hop) communication between the tags and the reader. Recent efforts addressed this limitation and enabled point-to-point communication among \emph{two} backscatter tags~\cite{karimi_rfid_2017,nikitin_rfid_2012,parks_sigcomm_2014}. 
	
\subsection{Beyond Two Tags: Multi-Hop Tag-to-Tag Networking}

An imminent and natural evolution step in the backscatter research area is to realize multi-hop tag-to-tag (T2T) networks. In these networks, \emph{low-complexity tags} are able to communicate with each other \emph{directly or through other tags} and readers provide only the carrier signal. Potential applications of such networks include (i) distributed T2T data collection (contacts exchange at a conference or warehouse stocktaking), and (ii) low-power distributed data processing. Apart from their applications, T2T multi-hop networks extend communication range of backscatter links~\cite[Sec. 5.1]{parks_sigcomm_2014}\footnote{We do not refer to near-field backscatter where short communication range is an asset, not a limitation.} without (i) increasing the transmission power of the readers or their receivers' sensitivity, or (ii) changing the positions of the readers or increasing their numbers. Moreover, direct T2T communication provides the following benefits: (i) the number of backscatter readers, and in turn (ii) their power-up cost\footnote{Impinj R1000 idle power consumption exceeds 24\,W~\cite[Table 1]{liu_jsac_2015}.}, as well as (iii) the cost of the overall infrastructure~\cite[Sec. I]{alevizos_arxiv_2017} are reduced---a signal generator has a very simple design compared to a backscatter reader. We speculate that future generations of T2T networks will also reduce communication delay: single hop T2T links will replace double hop \emph{reader-tag-reader} connections. 


\subsection{Multi-hop T2T: Research Questions and Contributions}
\label{sec:introduction-challenges}	

\begin{figure}
	\centering
	\subfloat[Standalone backscatter T2T tag]{\includegraphics[height=0.44\columnwidth]{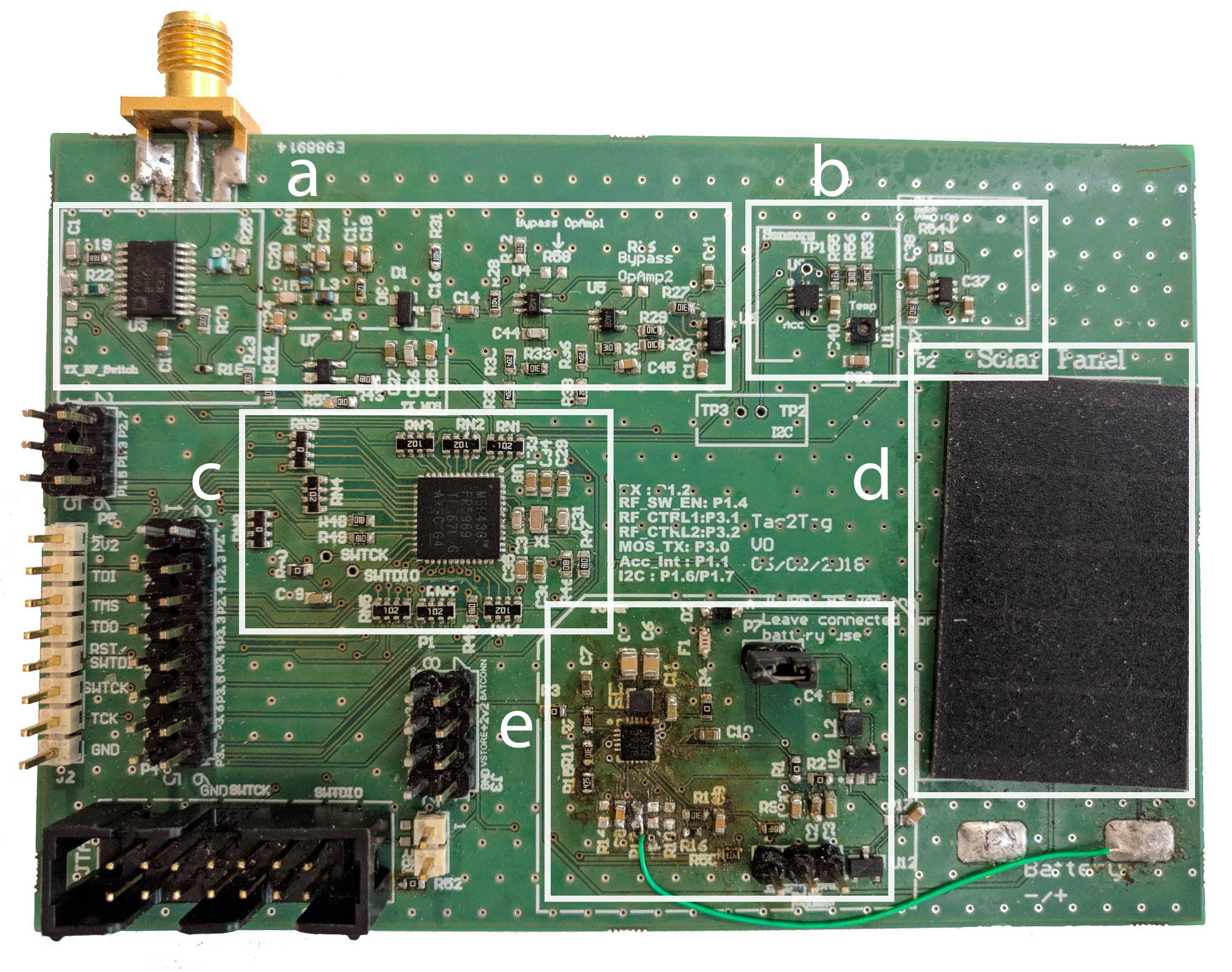}%
		\label{fig:fabricated_tag}}
	\subfloat[T2T transceiver]{\includegraphics[height=0.44\columnwidth]{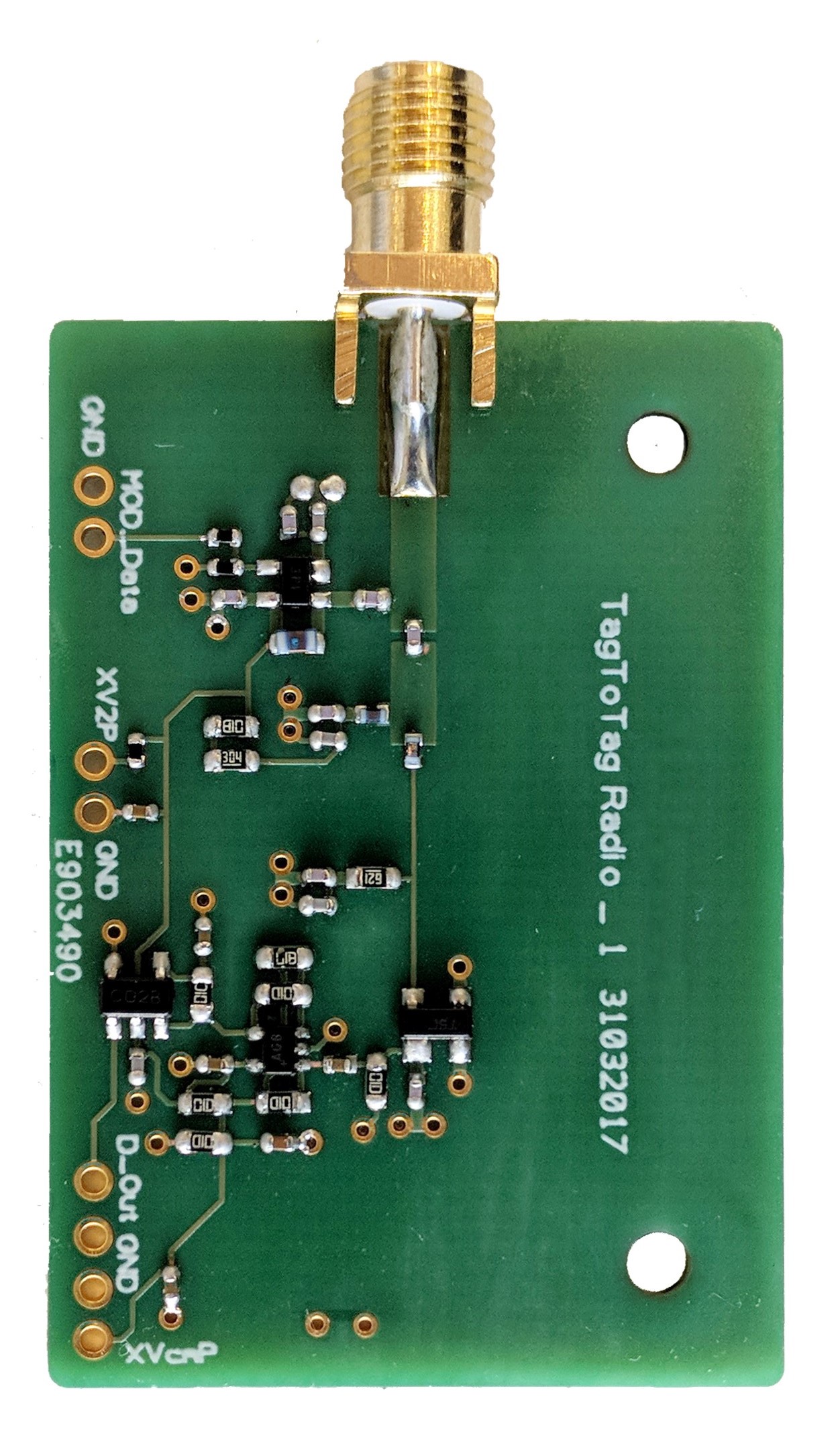}%
		\label{fig:backscatter_transceiver}}
	\caption{Two prototypes of backscatter T2T tag. Fig.~\ref{fig:fabricated_tag}: Integrated version (dimensions: 6.9\,cm$\times$10.1\,cm) with main sections marked as \textbf{a}: T2T transceiver, \textbf{b}: sensors, \textbf{c}: digital section with TI\,MSP430-family\,MCU~\cite{ti_msp430_website}, \textbf{d}: solar panel, \textbf{e}: power and energy harvesting. Fig.~\ref{fig:backscatter_transceiver}: T2T backscatter transceiver board only (dimensions: 3\,cm$\times$4.4\,cm) as an add-on for embedded platforms.}
	\label{fig:HW_pictures}
\end{figure}

Despite efforts in improving the transceiver design of (ambient) backscatter tags~\cite{parks_sigcomm_2014}, we are unaware of any prior work that is devoted to implement and demonstrate multi-hop backscatter communication. Therefore, our objective in this paper is to bring multi-hop T2T networks into reality. We start by raising the following question: \textbf{Q1}\emph{---how to create a network from tags, where tags communicate directly, using backscatter only?} This is immediately followed by \textbf{Q2}\emph{---what is the impact of multi-hop on bidirectional T2T communication range?} We conjecture that answering these questions is not trivial due to the complexity of simultaneous design of MAC, link layer, message passing and application layer in T2T networks. Our specific contributions that address our research questions are listed below:

\emph{Contribution 1}---\textbf{A complete Hardware/Software backscatter T2T ecosystem.} We demonstrate a fully operational backscatter tag-to-tag decode-and-relay multi-hop network. Our contribution includes (i) a novel backscatter tag hardware design, working in either energy passive or active mode (Fig.~\ref{fig:HW_pictures})---which we will open-source~\cite{tag-to-tag_source_files}---allowing the research community to build upon; (ii) in-depth characterization of T2T links, i.e. hop count, per link packet error rate; and as a consequence (iii) the first multi-tag backscatter T2T networking demonstration. 

\emph{Contribution 2}---\textbf{A novel Backscatter T2T MAC design.} Our network suite includes a novel MAC protocol. We show experimentally that backscatter reception is orders of magnitude more energy costly than transmission, contrary to active radios. Our MAC enables, per frame, transmission phase control; it uses  \emph{low-power listening} to conserve the energy of the microcontroller controlling the tag . We show that multi-hop increases the range of a T2T network significantly while providing the same level of robustness as the phase-shifting technique.


\emph{Contribution 3}---\textbf{The demonstration of T2T Network Bridging.} As a case study, we show that we can connect two T2T networks, served by separate exciters, by multi-hopping mechanism---tags that reside in the range of both exciters effectively serve as bridges between two networks.



\section{Backscatter T2T Network: Related Work}
\label{sec:related_work}

\subsection{Networking with T2T Backscatter Tags}
\label{sec:related_work_networking}

To the best of our knowledge, all studies in the domain of backscatter communication focused on the physical and/or the MAC layer aspects of the point-to-point communication between either the reader and tags, e.g.~\cite{talla_arxiv_2017,gummeson_mobisys_2012,wang_sigcomm_2012,zhang_mobisys_2012,yang_infocom_2017,zhang_sigcomm_2016,zhang_mobicom_2014} or between two tags only e.g.~\cite{liu_sigcomm_2013,shen_iotj_2016,karimi_rfid_2017,parks_sigcomm_2014,nikitin_rfid_2012,nikitin_rfid_2011}. For an overview of such backscatter architectures (including those that backscatter with FM, Bluetooth, or WiFi signals) see~\cite{talla_arxiv_2017}.

From the above studies, special attention needs to be put to the following works. Attempts, though unsuccessful, to design a protocol supporting multi-tag transmission were presented in~\cite{liu_sigcomm_2013}. A form of a backscatter-based network was presented in~\cite{wang_sigcomm_2012} (in the context of studies of induced collision resolution) and in~\cite{zhang_mobicom_2014} (in the context of studies of bit-by-bit MAC transfer). Unfortunately in both of these works the backscatter reader was still controlling the network and only single-hop communication is possible.

The closest work to ours is presented in~\cite{barnet_mobisys,bttn_iot}\footnote{We were aware of this work only at the final stage of the manuscript preparation.}. The authors present therein a working prototype of T2T network, however with no implementation of real message relaying~\cite[Sec. VI-B]{bttn_iot}, lacking full network stack implementation, based on one (line) topology, and implementing only one type of MAC based on frame repetition. 


\subsection{Backscatter T2T Tags Hardware}
\label{sec:related_work_hardware}

The first practical design of a backscatter T2T link was presented in~\cite{nikitin_rfid_2012,nikitin_rfid_2011}. Two important papers followed~\cite{liu_sigcomm_2013,parks_sigcomm_2014}, where the demodulation architecture based on an averaging filter (which we utilize in this paper) was introduced. An important work closest to ours is~\cite{karimi_rfid_2017} (see also associated work of~\cite{shen_iotj_2016,barnet_mobisys,bttn_iot}) where a similar transceiver architecture was proposed.
%
%
Another backscatter tag architecture worth mentioning is~\cite{liu_awpl_2011} that exploited multi-antenna diversity. Unfortunately, neither design details nor results of the signal demodulation from the tag have been presented therein. In~\cite{alevizos_arxiv_2017} a solar-powered tag was used (with a battery for energy collection during light absence). However, many details of their system, e.g. the detailed description of their tag design, are not reported. 
%



\section{Multi-hop Backscatter T2T Network Analysis}
\label{sec:analysis}



We now proceed with a set of theoretical results that show the benefit of multi-hop T2T networks. This analysis will provide fundamental insights which will be observed in a real T2T network implementation, presented in Section~\ref{sec:results}.

\subsection{Backscatter T2T Model Definition}
\label{sec:model_defnition}

Let us define a flat, square area of length $S_{a}$ with one exciter $E$ at position $(x_E, y_E)$ and $N$ static tags located at uniform random positions within the area. The exciter transmits a non-modulated signal of wavelength $\lambda_c$ and power $P_E$, which tags can backscatter to communicate with other tags with reflection coefficients $k_0, k_1$ for symbols 0, 1, respectively. The receiver sensitivity threshold for all tags is $P_s$ and the available power to backscatter at tag $n$ is~\cite[eq. (2)]{shen_iotj_2016}
\begin{equation}
P_n =  P_E G_E(\theta_{E,n}) G \lambda_c^2 \left(4\pi d_{E,n}\right)^{-2}
\label{eq:avail_power}
\end{equation}
where $d_{k,l}$ is the distance from device $k$ to $l$ (tag or exciter), $G_E(\theta_{E,n})$ is the antenna gain of the exciter in the angular direction to tag $n$, $\theta_{E,n}$, and $G$ the antenna gain of the tags. Then, the power received at tag $m$ from tag $n$ is~\cite[eq. (6)]{shen_iotj_2016}
\begin{equation}
P_{n,m} = P_n \left(k_0 G \lambda_c\right)^2 \left(4\pi d_{n,m}\right)^{-2}.
\label{eq:received_power}
\end{equation}
For simplicity, we consider only line of sight, free-space propagation with no fading or noise, as our goal is to provide the simplest bounds on network performance. As $k_0 < k_1$, (\ref{eq:received_power}) serves as a lower bound to the received power.

The T2T network is modeled as a weighted graph with a set of nodes $\{1, 2, \ldots, N\}$  and a set of directional links $\{L_{n,m}\}$. The weight of a link $L_{n,m}$ from tag $n$ to tag $m$ is defined as $L_{n,m}=P_{n,m}$ iff $P_{n,m}\geq P_s$ and $L_{n,m}=\varnothing$, otherwise. A single-hop T2T network is connected if $L_{n,m}\neq\varnothing, \forall n,m$. Conversely, a multi-hop T2T network is connected if there exists a path $\forall n,m$ node pairs.

\subsection{Analysis of Backscatter T2T Network}
\label{sec:analysis_subsection}

\subsubsection{Non-symmetric T2T Links}
\label{sec:non_symmetric_links}

A tag's available power depends on its distance from the exciter. This implies the following simple observation.



\begin{figure}
	\centering
	\includegraphics[width=0.85\columnwidth]{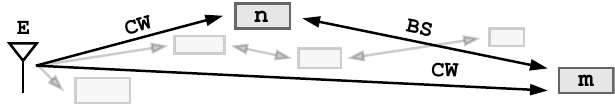}
	\caption{Illustration for received power analysis of T2T network; CW: carrier wave, E: exciter, BS: backscatter.}
	\label{fig:power_analysis}
\end{figure}

\begin{observation}
	\label{obs:non_symmetric}
	The ratio of received power between tags $n$ and $m$ in the forward link (from $n$ to $m$, Fig.~\ref{fig:power_analysis}) over the power received in the backward link (from $m$ to $n$) is quadratically proportional to $d_{E,m}/d_{E,n}$.
	%
	%
	\begin{IEEEproof}
		Directly from (\ref{eq:received_power}) we can write
		\begin{equation}
		\frac{P_{n,m}}{P_{m,n}} = \frac{G_E(\theta_{E,n})}{G_E(\theta_{E,m})} \left(\frac{ d_{E,m}}{d_{E,n}}\right)^2.
		\end{equation}
		This implies that in the useful case where tags are spread away from each other, each T2T link is non-symmetric, with much higher reception probability at the forward link (i.e. a link from the tag closest to the exciter to a tag further away), meaning that $P_{n,m}/P_{m,n}\gg1$.
	\end{IEEEproof}
\end{observation}

\subsubsection{Multi-hop Range}
\label{sec:multi_hop_range}

\begin{figure}
	\centering
	\includegraphics[width=0.85\columnwidth]{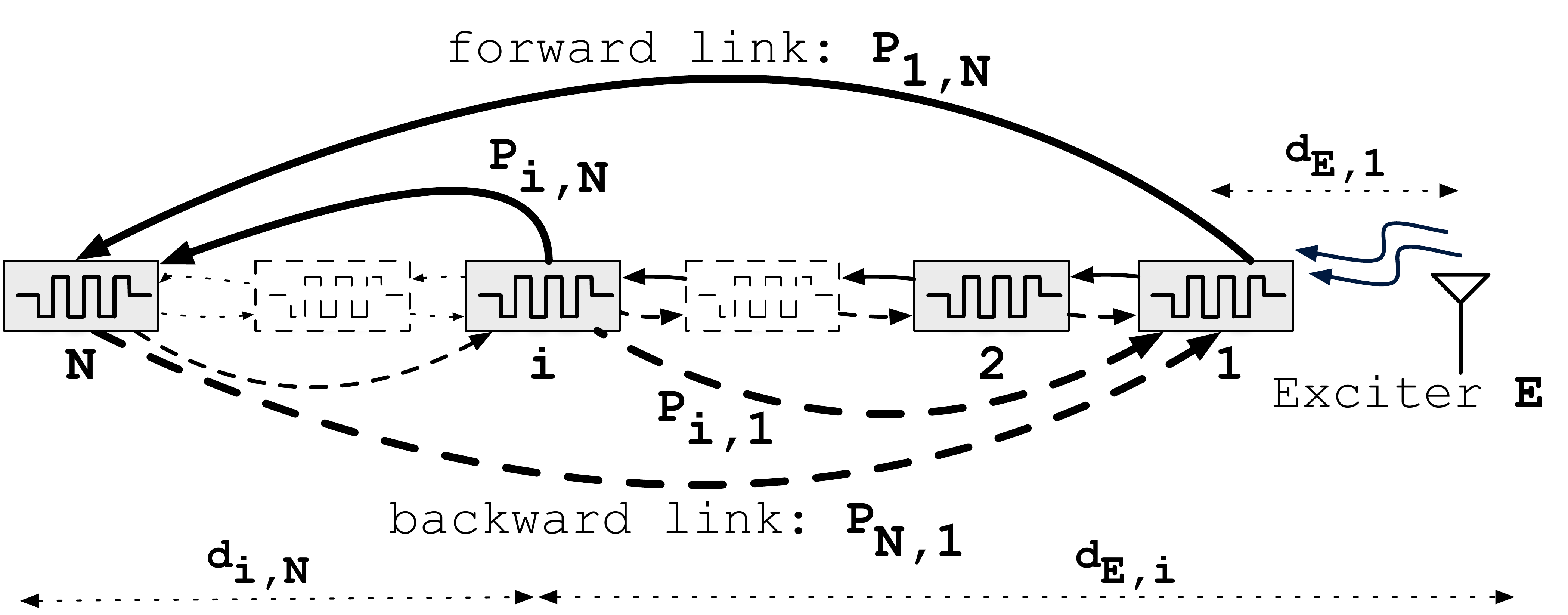}
	\caption{Multi-hop T2T received power analysis scenario.}
	\label{fig:MH_power_analysis}
\end{figure}

Next, let us analyze the range capabilities of T2T multi-hop networks in two ways: comparing received power with single-hop, and computing the maximum distance such networks can cover. 

\begin{observation}
	\label{obs:MH_gain}
		Considering tags 1 and $N$ (Fig.~\ref{fig:MH_power_analysis}), the received power from the transmitting tag using multi-hop is always greater than that using single-hop in the \emph{backward link} (from the most distant tag from the exciter to the tag closest to exciter).
	\begin{IEEEproof}
		Consider the linear topology depicted in Fig.~\ref{fig:MH_power_analysis}. With a little abuse of notation, let  $d_{E,1}=d_1$ and $d_{k-1,k}=d_k, \forall k\in[2,N]$. As shown in this figure, the received power in the backward link is $P_{N,1}$ for single-hop and $P_{i,1}$ for multi-hop, which depends on the tag $i$ over which the last hop is performed. For presentation compactness, define $\ell_a^b\triangleq\sum_{k=a}^{b}d_k$ that denotes the length of the path between tag $a$ and tag $b$ on the line topology. Using (\ref{eq:received_power}) again we can compute the power ratio
		%
		\begin{equation}
		\frac{P_{i,1}}{P_{N,1}}=\frac{P_i(d_{N,1})^{2}}{P_N(d_{i,1})^{2}} = \left(\frac{d_{E,N}d_{N,1}}{d_{E,i}d_{i,1}}\right)^2=
		\left(\frac{\ell_1^N \ell_2^N}{\ell_1^i \ell_2^i}\right)^2.
		\label{eq:bw_ratio}
		\end{equation}
		Note that~(\ref{eq:bw_ratio}) greater than one, as $i\in(1,N)$.
	\end{IEEEproof}
\end{observation}
\begin{corollary}
	Assume that tags are equally spaced on the line topology; i.e. $\forall k\in[2,N]: d_k=d_1=d$. Then, node $i$  that maximizes the received power in the backward link is $2$.
	\begin{IEEEproof}
		%
		For the backward link ($N$ to 1)
			\begin{equation}
			\operatorname*{arg\,max}_{i} \frac{P_{i,1}}{P_{N,1}}\Bigr|_{\substack{d_k=d}} = \operatorname*{arg\,max}_{i} \frac{\left(N\left(N-1\right)\right)^2}{\left(i\left(i-1\right)\right)^2} =2
			\label{eq:bw_ratio_same_d}
			\end{equation}
	\end{IEEEproof}
\end{corollary}
%


Let us now analyze the network's maximum range.
\begin{observation}
	\label{observation:max_range}
	Consider the multi-hop network of $N$ tags placed on a line. The optimal value of the distance between any two tags $i$ and $i-1$ that maximizes range and ensures communication in both ways is given by
	\begin{equation}
	d_i^* = \frac{1}{2}\left( \sqrt{ \left( \ell_1^{i-1}\right)^2 +4\epsilon}-  \ell_1^{i-1} \right)
	\label{eq:d_i}
	\end{equation}
	with $\epsilon\triangleq\lambda^2/(4\pi)^2Gk_0\sqrt{P_EG_E(\theta_{E,i})G/P_s}$.
	\begin{IEEEproof}
		As shown in Observation \ref{obs:non_symmetric}, the forward link (from $i-1$ to $i$) is less costly in power. Then, it suffices to guarantee communication in the backward link (from $i$ to $i-1$) by ensuring $P_{i,i-1}=P_s$ so that the received power is greater than the receiver sensitivity threshold. Therefore, it follows
		\begin{equation*}
		d_{E,i}d_{i-1,i}-\epsilon=d_i^2+\ell_1^{i-1}d_i-\epsilon=0,
		\end{equation*}
		from which $d_i^*=d_i$ can be solved.
	\end{IEEEproof}
\end{observation}



As more tags are added, they are placed closer together and eventually $d_i^*$ becomes small enough so that tags would be placed in the near field of their antenna. Therefore, we set a minimum inter-tag distance (equal to the Fraunhofer distance) as $d_\text{min} = d_{F} = 2D^2/\lambda \leq d_i^*$ with $D$ being the largest linear dimension of tag's antenna.

\subsubsection{Combating Phase Cancellation}
\label{sec:phase_cancellation}

Stemming from the fact that the energy source is dislocated from the T2T transmitter, it is possible that the backscatter signal and the un-modulated exciter signal arrive at a receiver with opposite phases and interfere destructively. This phenomenon is called \emph{phase cancellation}~\cite{shen_iotj_2016} and depends on the phase difference of arrival $\theta_{d, n}$: the difference of the phases of the signals arriving at T2T tag $n$ coming from the exciter and the transmitter. Phase cancellation effectively creates blind energy spots in the network and happens when~\cite[Eq. (18)]{shen_iotj_2016}
\begin{equation}
\theta_{d, n}= \theta_c = \cos^{-1}\left( -\frac{ (k_0+k_1) d_{E,m} \lambda_c G_n}{ 8 \pi d_{E,n} d_{n,m}} \right).
\end{equation}




The state-of-the-art solution to combat phase cancellation in T2T networks consists of sending every packet twice with a phase offset between them to ensure correct reception~\cite{shen_iotj_2016,bttn_iot}.
We will assess the network performance when phase shift technique is applied and when our multi-hop protocol is used. 


\paragraph{Analysis} \label{sec:analysis_efficiency}
Let us look at the average number of messages $E[m]$ and average transmission time $E[t]$ it takes to deliver a frame for any source-destination T2T pair, as well as the probability of successfully doing so, $\Pr(s)$. Then, let us consider two network typologies: (i) a \emph{known network topology} in which all nodes have complete information of the network connections, and (ii) an \emph{unknown topology}. These two cases are, in turn, split into cases where the source and the destination tags are within single-hop range or not (denoted as `SHR' and `MHR', respectively).

Table~\ref{table:efficiency_analysis} collects all combinations for cases A and B. Case A is the single-hop phase shifting solution---repeating the same message twice with 90$^{\circ}$-shifted phase, as proposed in~\cite{shen_iotj_2016,bttn_iot} and case B is the multi-hop protocol. When network topology is known the optimum path can be computed, which also enables avoiding links that are down due to phase cancellation effect (which we assume to happen with probability $p_c$) by switching the phase of the transmitted frame. Furthermore, the minimum hop count, $H$, is reached and $E[t]$ is minimized, which is composed of the time-of-flight of a frame, $t_f$, and the processing time taken by a node to forward a frame, $t_p$. In the case of unknown topology, we assume an average number of relay nodes $M$ which are able to forward the message and reach the destination.

\paragraph{Result} Comparing cases A and B, Table~\ref{table:efficiency_analysis}, we see that our solution, i.e. Case B, improves network utilization or communication success probability in most cases. However, if source and destination happen to be in range at unknown topology, it is possible that the usage of the T2T network is increased and/or the success rate is reduced depending on $M$ and $p_c$.

\begin{table}
	\scriptsize
	\centering
	\caption{Backscatter T2T Network Efficiency Analysis Results; `MHR': multi-hop range, `SHR': single-hop range}
	\label{table:efficiency_analysis}
	\begin{tabular}{cc|l|l|}
		\cline{3-4}
		\multicolumn{1}{l}{}           & \multicolumn{1}{l|}{} & \multicolumn{1}{l|}{Case A: \emph{phase shifting}} & \multicolumn{1}{l|}{Case B: \emph{multi-hop flooding}} \\ \hline
		\multicolumn{1}{|c|}{}         &                       & $E[m]=2$                           & $E[m]=H$                           \\
		\multicolumn{1}{|c|}{}         & MHR                    & $E[t]=2t_f$                           & $E[t]=H(t_f+t_p)-t_p$                           \\
		\multicolumn{1}{|c|}{Known}    &                       & $\Pr(s)=0$                           & $\Pr(s)=1$                           \\ \cline{2-4} 
		\multicolumn{1}{|c|}{topology} &                       & $E[m]=2$                           & $E[m]=1$                           \\
		\multicolumn{1}{|c|}{}         & SHR                    & $E[t]=2t_f$                           & $E[t]=t_f$                           \\
		\multicolumn{1}{|c|}{}         &                       & $\Pr(s)=1$                           & $\Pr(s)=1$                           \\ \hline
		\multicolumn{1}{|c|}{}         &                       & $E[m]=2$                           & $E[m]=M+1$                           \\
		\multicolumn{1}{|c|}{}         & MHR                    & $E[t]=2t_f$                           & $E[t]=M\left(t_f+t_p\right)+t_f$                           \\
		\multicolumn{1}{|c|}{Unknown}  &                       & $\Pr(s)=0$                           & $\Pr(s)=1-p_c^{M+1}$                           \\ \cline{2-4} 
		\multicolumn{1}{|c|}{topology} &                       & $E[m]=2$                           & $E[m]=M+1$                           \\
		\multicolumn{1}{|c|}{}         & SHR                    & $E[t]=2t_f$                           & $E[t]=M\left(t_f+t_p\right)+t_f$                           \\
		\multicolumn{1}{|c|}{}         &                       & $\Pr(s)=1$                           & $\Pr(s)=1-p_c^{M+1}$                           \\ \hline
	\end{tabular}
\end{table}


\begin{table}
	\begin{threeparttable}
		\scriptsize
		\caption{Realistic model of T2T network used in numerical example}
		\label{table:parameter_definitions}
		\centering
		\begin{tabular}{ | r | l | c | l | }
			\hline
			Symbol & Value & Units & Description \\ 
			\hline\hline
			$f_c$ & 868 & MHz & Center frequency \\
			$(k_0; k_1)$ & (0.4; 0.9) & --- & Reflection coefficients for 0 and for 1\\
			$(P_E$; $P_s)$ & (33; -50) & dBm & Output power of exciter, sensitivity of tags \\
			$(G; G_E)$ & (0; 4) & dBi & Antenna gain: tags; exciter \\
			$(\theta_E$; $\theta_c)$ & (-45; 40) & $\degree$ & Exciter antenna beam direction; beam width \\
			$(x_E, y_E)$ & (0; 3) & m & Cartesian coordinates of exciter \\
			$S_{a}$ & 30 & m & Side of the square area of the T2T network \\
			$d_1$ & 3 & m & Distance from exciter to tag 1 (in Fig.~\ref{fig:MH_range}) \\
			\hline
		\end{tabular}
	\end{threeparttable}
\end{table}

\subsection{Backscatter T2T Network: Numerical Results}
\label{sec:analysis_numerical_results}

\begin{figure}
	\subfloat[T2T multi-hop coverage]{\includegraphics[width=0.5\columnwidth]{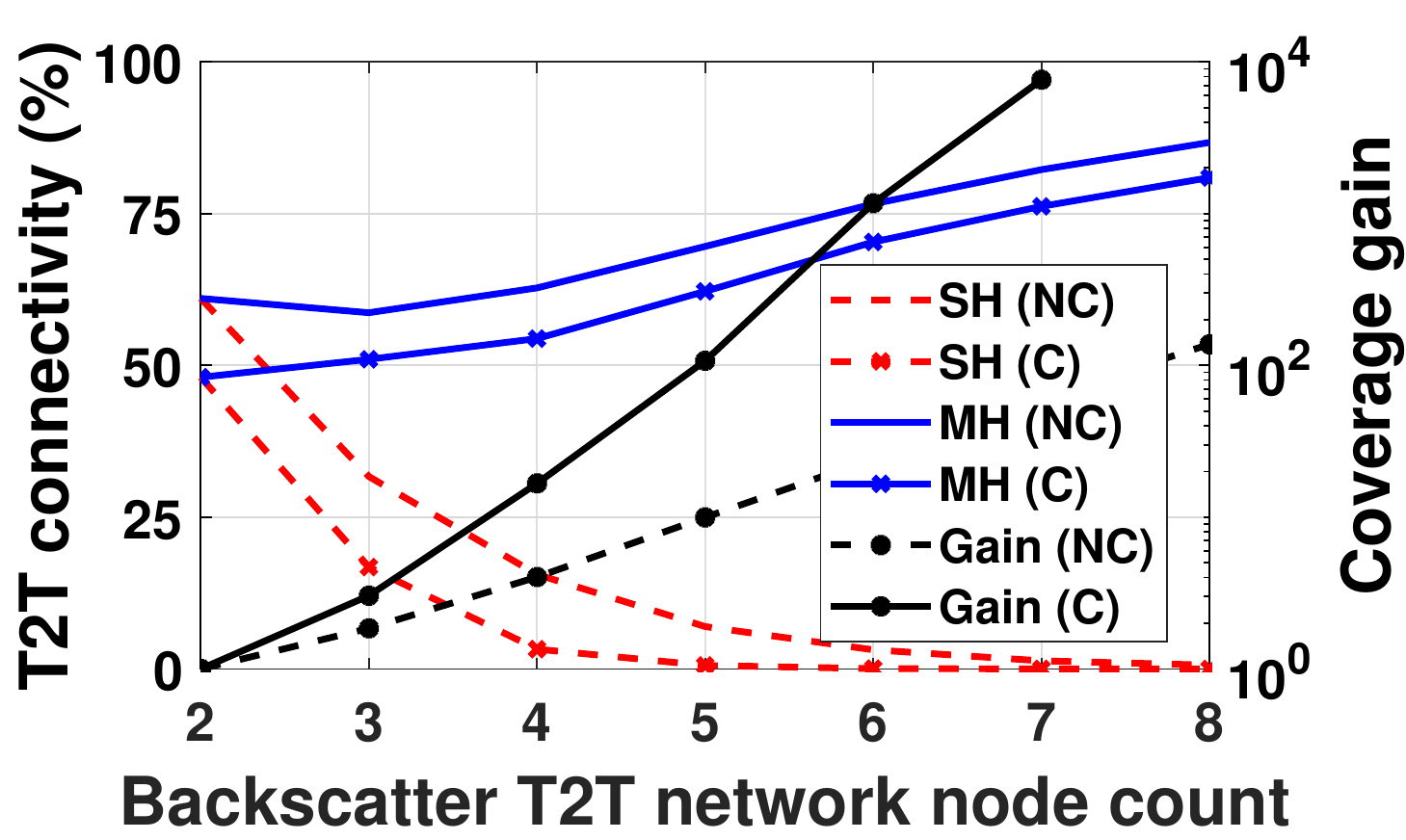}%
		\label{fig:cancel_vs_no_cancel_coverage_sas3_33dBm}}
	\subfloat[T2T maximum multi-hop range]{\includegraphics[width=0.5\columnwidth]{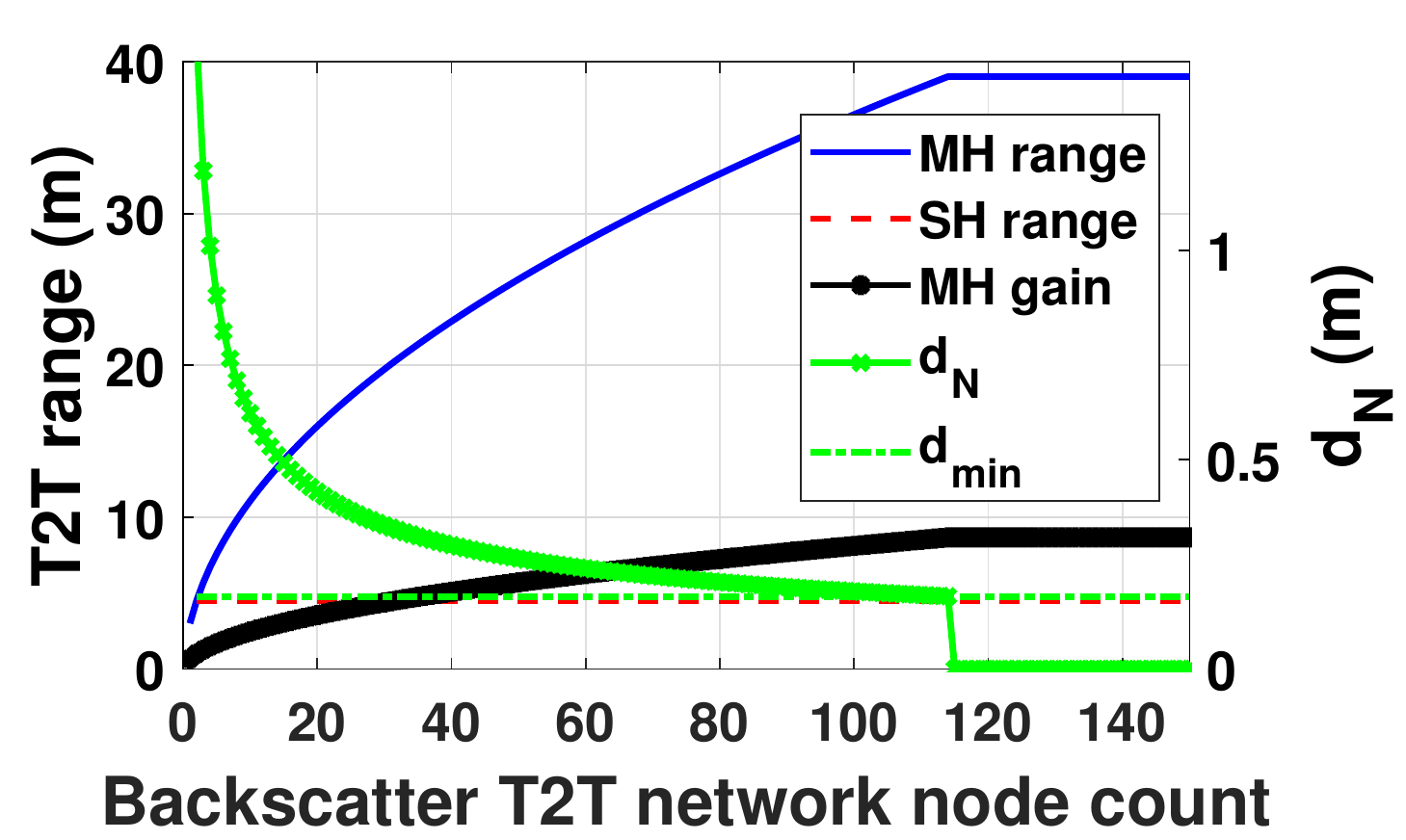}%
		\label{fig:MH_range}}
	\caption{Backscatter T2T network numerical example with simulation parameters listed in Table~\ref{table:parameter_definitions}. Fig.~\ref{fig:cancel_vs_no_cancel_coverage_sas3_33dBm}: multi-hop versus single-hop full connectivity (probability that \emph{all} nodes are connected with each other, either only directly, SH, or through hop, MH) with and without phase cancellation; Fig.~\ref{fig:MH_range}: multi-hop maximum range in one-dimensional T2T network; \emph{NC}: no phase cancellation, \emph{C}: phase cancellation, \emph{MH}: multi-hop, \emph{SH}: single-hop.}
	\label{fig:analysis_sims}
\end{figure}

We illustrate core analytical results with a numerical example. We simulate an instance of a backscatter T2T network with the parameters given in Table~\ref{table:parameter_definitions}. Simulation code~\cite{tag-to-tag_source_files} is written in Matlab. We generate each instance of a T2T network by randomly placing nodes on a square of side $S_a$ and checking connectivity at every iteration. Each simulation point is an average of 10000 runs. The results are given in Fig.~\ref{fig:analysis_sims}.

\textbf{Result 1}---\emph{Multi-hop coverage.} We define coverage when \emph{all} T2T tags are connected with each other. Backscatter T2T single-hop coverage does rapidly decrease with $N$, see Fig.~\ref{fig:cancel_vs_no_cancel_coverage_sas3_33dBm}, and more so when accounting for phase cancellation, improving the multi-hop gain. Also, the phase cancellation effect becomes less relevant with a denser T2T network. 


\textbf{Result 2}---\emph{Maximum multi-hop range.} Fig.~\ref{fig:MH_range} presents the maximum communication range as a function of the number of nodes achieved by a one-dimensional backscatter T2T network while ensuring two-way communication. The maximum range was computed according to the analysis presented in Observation~\ref{observation:max_range}. When $d_N<d_{\min}$, T2T network range can no longer be increased, but until then, multi-hop range increases logarithmically, opposite to inter-tag distance $d_N$.

\textbf{Remark:}---\emph{Limitation of T2T network.} The main bottleneck of T2T network is the small coverage, caused by the low signal detection threshold of the tags, $P_s$. Precisely because of this, traditional tag-to-reader networks would outperform a T2T counterpart in communication range.


\section{Backscatter T2T Tag Design: Hardware}
\label{sec:tag}

\begin{figure}
	\centering
	\includegraphics[trim=0 10 0 0,width=.85\columnwidth]{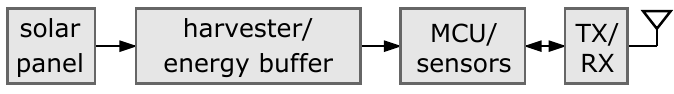}
	\caption{Backscatter T2T tag block diagram.}
	\label{fig:system_overview}
\end{figure}

We proceed with the description of backscatter T2T tag architecture, see Fig.~\ref{fig:system_overview}. This section will introduce the T2T tag hardware, while Section~\ref{sec:network} will introduce the supporting carrier generator and design choices on the tags deployment.

\begin{figure}
	\centering
	\includegraphics[trim=0 355 0 0,clip,width=\columnwidth]{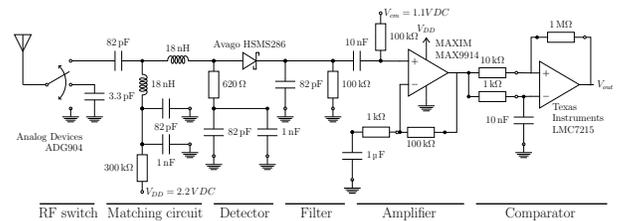}
	\caption{Backscatter T2T transceiver; its implementation is shown in Fig.~\ref{fig:backscatter_transceiver}.}
	\label{fig:tag_schematics}
\end{figure}

\subsection{Backscatter T2T Tag: Backscatter Transceiver}
\label{sec:tag-transciever}

Our T2T backscatter transceiver is designed with energy consumption reduction in mind. It, therefore, avoids all energy-hungry component such as ADCs or multi-stage power amplifiers. Its complete hardware design, introducing novel adaptations with respect to~\cite[Fig. 6]{liu_sigcomm_2013},~\cite[Fig. 4]{parks_sigcomm_2014},~\cite[Fig. 3]{karimi_rfid_2017} is presented in Fig.~\ref{fig:tag_schematics} and discussed stage-by-stage below. We note that to design and simulate the transceiver's (i) RF (including on-PCB antenna matching circuit) and (ii) the low-frequency part (including filter and amplifier parameters), ADS version 2012~\cite{ads_website} and OrCAD Capture version 17~\cite{orcad_website} software suites were used.

\subsubsection{Antenna Matching} The transceiver is preceded by 50\,$\Omega$ antenna matching circuit tuned at 868\,MHz center frequency, i.e. working within 863--870\,MHz (SRD860) band. Matching is built with a transmission line and passive SMD components.

\subsubsection{Backscatter Transmitter} It is composed of an Analog Devices ADG904 SP4T RF switch inducing three modulation states (reflecting, reflecting with 90$^{\circ}$phase shift, and non-reflecting an impinged radio frequency signal from external transmitter: non-dedicated as in~\cite{parks_sigcomm_2014,yang_infocom_2017}, or dedicated). The phase shift enables the tag to combat the phase cancellation problem of a T2T link~\cite{shen_iotj_2016} discussed also in Section~\ref{sec:phase_cancellation}. The phase shift  technique implementation is based on a design proposed in~\cite[Sec. IV-A]{shen_iotj_2016}, i.e. phase change-inducing switch connected in-between antenna and the SMA port of the transceiver to reflect the same bit with an inverted phase. The embedded version of the tag, Fig.~\ref{fig:fabricated_tag}, uses an SMD version of the switch, while transceiver-only version, Fig.~\ref{fig:backscatter_transceiver}, used ADG904 RF SP4T switch evaluation board connected via the antenna port (not shown on the photograph). The RF switch is controlled digitally by an embedded microcontroller, which is described in Section~\ref{sec:tag-computing}. 

\subsubsection{Backscatter Receiver} It is composed of (i) an envelope detector (to filter out the mix of carrier and other tag's backscatter signal), (ii) a low-pass/high-pass filtering block, (iii) a baseband amplifier, and (iv) a Texas
Instruments LMC7215 comparator (i.e. a one-bit ADC). 

\paragraph{Envelope Detector} To filter out the RF components of the carrier, a rectifier diode (Avago HSMS286) followed by a parallel RC network are used. We enhance the voltage swing of the detector input RF signal by biasing the diode to overcome its threshold voltage. This distinguishes our design from the envelope detector introduced in~\cite[Fig. 4]{karimi_rfid_2017} which uses voltage doubler. The benefit of biasing is a lower junction resistance of the diode, which enables easier matching to a 50\,$\Omega$­ transmission line~\cite[p. 6]{hp_note_963}. A shunt resistor is added before the diode for improved matching (reducing reflection losses) for a desirable $S_{11}<-10$\,dB. For our design $S_{11}=-32.31$\,dB. This result was obtained with ADS.

\paragraph{High-Pass Filter} The output of the envelope detector is passed to a high-pass filter to block the DC component (since there is information in it) of the carrier wave.

\paragraph{Ultra-Low-Power Amplifier} We propose to amplify the output of the high-pass filter\footnote{This idea was independently proposed in~\cite[Sec. III-B]{karimi_rfid_2017}.}. Although diode is biased to improve the baseband signal swing, this voltage swing is still in $\mu$V range at low power levels. If the comparator would be directly connected to the envelope detector, such as in~\cite[Fig. 6]{liu_sigcomm_2013},~\cite[Fig. 4]{parks_sigcomm_2014}, the total sensitivity would be limited due to the comparator offset of typically few millivolts. In our design we use off-the-shelf amplifier, Maxim Integrated MAX9914, (Fig.~\ref{fig:tag_schematics}), instead of a custom-build one like in~\cite[Sec. III]{karimi_rfid_2017}. This way, we keep the monetary costs low, enabling widespread use of our platform among the research community~\cite{tag-to-tag_source_files}\footnote{We note that up to now regrettably \emph{no tag-to-tag hardware}, including~\cite{parks_sigcomm_2014,liu_sigcomm_2013,yang_infocom_2017,karimi_rfid_2017} \emph{has been open-sourced}.}. Use of an amplifier is being traded for increased energy consumption compared to existing state-of-the-art tags---nevertheless, is still many times lower than low-power active radios~\cite[Fig. 3]{talla_arxiv_2017}. A gain of 100 was chosen for amplification.
%

\paragraph{Comparator} The output of the amplifier is fed into the inputs of the comparator. The comparator's inverting terminal includes an averaging circuit to determine the right threshold for bits detection\footnote{Potential improvements such as dynamic tracking of a threshold for comparator and self-interference cancellation circuit advocated in~\cite[Sec. 2.2]{zhang_sigcomm_2016} were left out for future design.}. Finally, to minimize the number of false triggers due to noise, a hysteresis of 10\,mV is added. The used comparator was Texas Instruments LMC7215.


\subsection{Backscatter T2T Tag: Power Supply}
\label{sec:tag-supply}

The communication distance of a passive backscatter tag, i.e. powered by the RF energy of a carrier generator, is limited by the minimum power that needs to be supplied at the RF harvester (which should be greater than {$\text{-25}$\,dBm}~\cite[Sec. 2]{griffin_apm_2009}). Therefore, to extend  tags communication range, and to simplify the design, we choose for semi-passive (energy-assisted) architecture~\cite[Sec. II]{alevizos_arxiv_2017} by powering tags from the non-RF energy source, i.e. a solar panel.

Specifically, to power the transceiver and the microcontroller of our tags, two monocrystalline 2\,V/44.6\,mA IXOLAR SLMD121H04L solar panels ~\cite{SLMD121H04L_website} were connected in series to TI BQ25570EVM-206 evaluation board~\cite{BQ25570EVM-206_website} (in case of transceiver design only) hosting TI BQ25570~\cite{BQ25570_website} harvester power management circuit.
%
%
Finally, we note that the parameters of the regulating nano-power management circuit of TI BQ25570 were set following TI BQ25505/70 Design Help V1.3 tool~\cite{BQ25570_website}. The energy is stored in a 470\,$\mu$F super-capacitor, while for prototyping and measurements tags were powered directly via the USB interface of the microcontroller or via a battery. 


\subsection{Backscatter T2T Tag: Computing Engine}
\label{sec:tag-computing}

The backscatter transceiver is connected to TI MSP430FR5969 MCU, i.e. with non-volatile memory. Microcontroller's role is to encode and decode incoming frames of the protocol introduced in Section~\ref{sec:protocol}. In the case of transceiver-only fabrication of the T2T tag (Fig.~\ref{fig:backscatter_transceiver}), the transceiver is connected to a launchpad~\cite{EXP430FR5969_website} (for ease of prototyping), in contrary to the self-contained fabricated tag (Fig.~\ref{fig:fabricated_tag}).

\subsection{Backscatter T2T Tag: Complete Design}
\label{sec:tag-complete}

We have fabricated two types of T2T tags, both presented in Fig.~\ref{fig:HW_pictures}. The first one is a transceiver board only which can be connected to any microcontroller, energy harvester or power supply of choice (Fig.~\ref{fig:backscatter_transceiver}). All components of this transceiver were hand-soldered on a two-layer 3\,cm$\times$4.3\,cm printed circuit board. The second design is all-in-one tag (Fig.~\ref{fig:fabricated_tag}), containing all required components on a single PCB. Both types of fabricated tags used 868\,MHz 3\,dBi gain 50\,$\Omega$ omni-directional whip antenna connected through an SMA connector to the input port. In the rest of the paper, we have experimented only with the first option due to cost considerations. The total cost of the T2T tag transceiver, based on standard hardware suppliers, did not exceed 25\,\euro. 





\section{Backscatter T2T Tag Design: Tags and Carrier Generators}
\label{sec:network}

\emph{T2T Tag Population.} We locate our T2T network indoors. Tags can be placed anywhere---either in line-of-sight to the exciter or not. Each T2T tag has the same level of hierarchy, master-master, in contrast to architecture of~\cite{nikitin_rfid_2012}. 



\emph{Number of Carrier Generators.} Favoring mono-static setup against complex/costly multi-static one~\cite{alevizos_arxiv_2017} we use $N_g=1$ carrier generators (with no connectivity options either with other T2T tags or the outside world). Naturally, in any T2T network $N\gg E$ is desired. 

\emph{Carrier Generator.} In our experiments, we used two carrier generators: Agilent E4438C ESG and Hewlett-Packard HP8648C (use of which specific generator will be reported explicitly in the respective experimental results in Section~\ref{sec:results})\footnote{We note that although our tags were designed to work with ambient carrier signals, due to extremely weak signal levels and lack of control over the location of the carrier wave, we have tested our network with signal sources we had full control of.}. Any signal generator is connected to Liard S928PCR 902--928\,MHz 8\,dBic, right-hand circularly polarized antenna. The maximum output power of the generator is set to 20\,dBm (for comparison, a 13\,dBm output power carrier generator was used in~\cite{alevizos_arxiv_2017}). We do not perform any duty cycling, which is allowed by the regulatory bodies for some RFID spectrum ranges, i.e. 915--921\,MHz~\cite[Sec. 4.3.7.3]{etsi_standard_302_208}. 


\section{Backscatter T2T Tag Design: Protocol Suite}
\label{sec:protocol}

We proceed with the introduction of our protocol suite of the backscatter T2T network. In particular, we introduce a novel MAC design that perfectly fits with the energy constraints of the T2T tags, followed by the link and application layer.

\subsection{Backscatter T2T Medium Access Control}
\label{sec:t2t_mac}


\begin{table}
	\scriptsize
	\centering
		\begin{threeparttable}
		\caption{Energy consumption of T2T Tag; Transceiver (as in Fig.~\ref{fig:backscatter_transceiver}) connected to MSP430 evaluation board~\cite{EXP430FR5969_website}\tnote{$\dagger$}}
			\begin{tabular}{| c | c |}
				\hline
				Operation mode & Energy consumption (mW) \\
				\hline\hline
				Reception only & 1.3 \\
				Transmission only\tnote{$\star$} & 0.7 \\
				MCU only (Low Power Mode 0)\tnote{$\ddagger$}~\cite{ti_msp430_website} & 2.2 \\
				\hline
			\end{tabular}
			\label{tab:power_consumption}
			\begin{tablenotes}
				\scriptsize
				\item[$\star$] T2T transmitter set to transmit continuous stream of bits
				\item[$\dagger$] Measurements using~\cite{monsoon}, powered at 3.6\,V; 16\,MHz MCU clock cycle
				\item[$\ddagger$] Includes powering all on-board peripherals, including USB controller
			\end{tablenotes}
		\end{threeparttable}
\end{table}






The MAC layer of the studies \cite{bttn_iot,liu_sigcomm_2013} implements a continuous carrier sensing mechanism to detect incoming bits as well as to avoid interference between neighboring tag-to-tag links. This approach demands the receiver and the MCU to be always kept on in order to receive incoming packets---leading to considerable energy overhead. 

\textit{Observation 1}---\textbf{Backscatter reception is more costly.} Backscatter transmission is cheap, since it is performed only by toggling the MCU port connected to the RF switch (see Fig.~\ref{fig:tag_schematics} or~\cite[Fig. 3]{talla_arxiv_2017}). However, contrary to active radios\footnote{Refer to, e.g., LTE Cat NB1 radio~\cite[Table 4.2.3]{ublox_saran2}, where transmission expenditure increases with transmission output power.}, backscatter reception is far more energy costly as compared to backscatter transmission. This is due to the additional energy cost of the receiver circuitry and the uncontrollable MCU false interrupts during the reception window. This is proven by our example T2T tag energy consumption measurements, see Table~\ref{tab:power_consumption}: reception is almost \emph{twice as expensive} as transmission. All in all, the MCU is the most power-hungry component of the tag \emph{requiring sleep scheduling}.


\textit{Observation 2}---\textbf{Ambient noise prevents energy-efficient wake-up radio operation.} Since the comparator circuit triggers MCU at each bit transition, the backscatter receiver can be seen as a \emph{wake-up} radio, which eliminates \emph{idle listening}~\cite{chen2013range}. However, due to ambient noise, the MCU can also be triggered in the absence of any backscatter transmission. These \emph{false triggers} are dependent on the selection of the comparator threshold and will wake-up the MCU from the sleeping state quite frequently---leading to wasted energy due to false wake-up and listen periods.

\textit{Observation 3}---\textbf{Packet detection using carrier sense is inefficient for low data rates.} Employing carrier sense to detect if there is an incoming packet, as e.g. in~\cite[Sec. 4.2.1]{liu_sigcomm_2013}, requires the backscatter receiver to be \emph{always} kept on in order to process incoming bits. However, this leads to significant overhead due to the energy cost of the reception  (\emph{vide} Observation 1) and false triggers (\emph{vide} Observation 2). 

\textit{Observation 4}---\textbf{Synchronous operations are infeasible.} Introducing synchronous protocols; e.g. forcing tags to wake-up their backscatter receivers just before their assigned slots as in TDMA, is not feasible as of now due to the reasons being~\cite[Sec. 3]{aantjes_hlpc_2016}: (i) backscatter tags would not have timekeeping mechanism with frequent power losses; (ii) there is no central entity in the network which is continuously powered, assigning transmission schedules (such as RFID reader)---making approaches such as ~\cite[Sec. 3.5]{talla_arxiv_2017} inapplicable. 

\begin{figure}
	\centering	
	\subfloat[Low power listening]{\includegraphics[width=0.49\columnwidth]{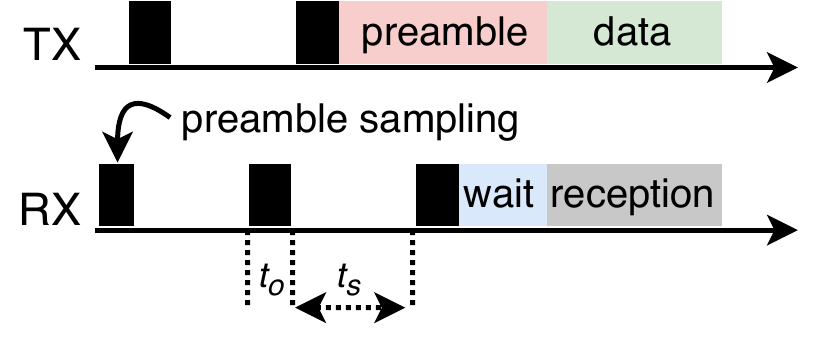}\label{fig:lpl-mac}}
	\subfloat[T2T MAC finite state machine]{\includegraphics[width=0.49\columnwidth]{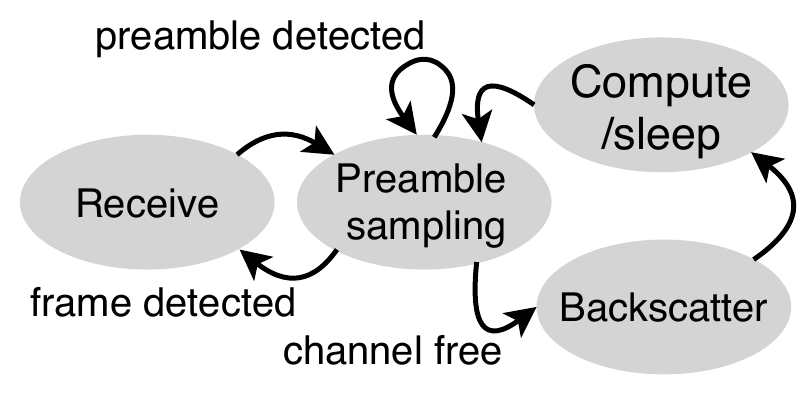}\label{figure:mac-state-machine}}
	\caption{Schematic representation of the proposed MAC protocol. Fig.~\ref{fig:lpl-mac}: low-power listening diagram (for $t_s$ and $t_o$ see Table~\ref{tab:mac_parameters}); Fig.~\ref{figure:mac-state-machine}: MAC state machine (for both transmitter and receiver).}
\end{figure}

\emph{Proposition}---\textbf{MAC Paradigm for Backscatter Tags: Low-Power Listening} Given \emph{Observations 1--4} we advocate for \emph{asynchronous} MAC design based on \emph{low power listening} inspired by~\cite[Sec. 3]{polastre_sensys_2004}.

\begin{table}
	\centering
	\scriptsize
	\begin{threeparttable}
		\caption{Selected T2T MAC parameters; see~\cite{tag-to-tag_source_files} for details}
		\label{tab:mac_parameters}
		\begin{tabular}{|c|c|c|c|}
			\hline
			Parameter & Description & Value & Unit \\
			\hline\hline
			$b_r$, $b_t$ & RX and TX buffer size  & 8, 8 & Frames \\
			$l_b$ & Bit length & Section~\ref{sec:mac_frame} & CC\tnote{$\star$} \\ 
			$t_s$ & Sleep period & 26.5 & ms \\ 
			$t_x$ & Frame reception timeout & 15 & ms \\ 
			$t_p$ & Frame preamble length & 36 & ms \\ 
			$t_s$, $t_g$ & Inter-frame time & 0.25 & ms \\ 
			$t_{l}$, $t_{h}$ & Timer jitter low and high & 25, 37.5 & $\mu$s \\ 
			$t_o$ & Channel observation period & 6.1 & ms \\ 
			$n_{o}$ & Min bit transitions within $t_{o}$ & 8 & Bits \\ 
			\hline 
		\end{tabular}
		\begin{tablenotes}
			\item[$\star$] \scriptsize {CC: cycles of SMCLK clock of MCU~\cite{EXP430FR5969_website}  at 16\,MHz}
		\end{tablenotes}
	\end{threeparttable}
\end{table}



\subsubsection{Low Power Listening and MAC State Machine}
\label{sec:mac_state_machine}

High-level design of our MAC is illustrated in Fig.~\ref{fig:lpl-mac}, while MAC finite state machine is presented in Fig.~\ref{figure:mac-state-machine}. Briefly, a T2T node wakes up periodically every 26.5\,ms (see Table~\ref{tab:mac_parameters}), detects the preamble (by reading bits at the comparator output port and searching for the predefined preamble, Table~\ref{tab:mac_parameters}) and synchronizes to the delimiter of the received frame. A frame validation is started by calculating the CRC of the received frame and comparing it to the CRC bytes of the frame. If they match the freshness of the frame is checked by searching a ring buffer that keeps track of the most recent $y$ (in our implementation $y=10$) frames. If the frame is new and the frame \texttt{Receiver ID} equals the node \texttt{ID}, then the payload is saved in the memory and made accessible to the application layer. However, if the \texttt{ID}s mismatch the frame will be saved in the transmission buffer to be forwarded.

In order to avoid collision each data transmission is preceded by channel observation (preamble sampling) and a long preamble (see Table~\ref{tab:mac_parameters}) is transmitted to wake up surrounding nodes. Once a frame is received or discarded the MCU transitions to the low-power mode (i.e. Low Power Mode 0 of the MSP430 MCU used~\cite{ti_msp430_website}) turning off the T2T backscatter receiver until the next wake-up. Furthermore, the MAC cycle is randomized to prevent nodes from waking up and estimating the channel at the same time, which may lead to concurrent transmission, see Fig.~\ref{figure:mac-state-machine}. Our MAC enables nodes to transmit with and without a phase shift, of 90$^{\circ}$, to enable nodes to cope with the dead spots in backscatter networks~\cite{shen_iotj_2016}. 


\subsubsection{MAC Frame Structure}
\label{sec:mac_frame}

We implement a frame structure described in Table~\ref{tab:frame_structure}. Bits in the frame are coded with FM0 modulation. Finally, we note that bit lengths can vary in a set [1600, 16000, 31250] of SMCLK clock cycles that corresponds to respective [10\,k, 1\,k, 512]\,bps data rates. In the actual implementation, we have chosen for the highest clock rate/bit rate, i.e. 1600 clock cycles/10\,kbps, respectively. 

\begin{table}
	\centering
	\scriptsize
	\begin{threeparttable}
		\caption{Frame Structure of the Implemented T2T MAC Protocol}
		\label{tab:frame_structure}		
		\begin{tabular}{|c|c|c|}
			\hline
			Field & Length (B) & Notes \\
			\hline\hline
			Preamble & Time based & \texttt{0xBB}\tnote{$\star$} \\
			Start Frame Delimiter & 1 & \texttt{0xAA}\tnote{$\star$} \\
			Sender ID & 1 & Node ID \\
			Receiver ID & 1 & Node ID \\
			Message Type\tnote{$\dagger$} & 1 & \texttt{Broadcast} ID Byte = \texttt{0xFF} \\
			Message ID & 1 & --- \\
			Payload	& 4 & --- \\
			CRC & 2 & CRC-CCITT hardware-based~\cite{EXP430FR5969_website}\\
			\hline
		\end{tabular}
	\begin{tablenotes}
		\item[$\star$] \scriptsize {Values obtained experimentally}
		\item[$\dagger$] \scriptsize {In experiments only \texttt{Broadcast} used; other types like \texttt{ACK}, \texttt{Beacon}, \texttt{Data} possible}
	\end{tablenotes}
\end{threeparttable}
\end{table}







\subsubsection{T2T Tag Addressing} All tags have a predefined addresses of 1\,B long (Table~\ref{tab:frame_structure}). We do not implement any tag discovery protocol at this stage.

\subsection{Backscatter T2T Link Layer}
\label{sec:t2t_link_app}


To increase the robustness of the T2T network we chose to implement a flooding mechanism to forward the messages. Flooding also implicitly eliminates the need for implementing error correcting mechanism, since multiple copies of a frame are forwarded. To limit the number of the messages being forwarded each node is allowed to backscatter a new frame $z$ times (in our implementation $z=1$). Furthermore, to reduce the probability of collisions each tag observes the channel before backscattering and its wake-up cycles are randomized within a range of [0, 5]\,ms.



\subsection{Backscatter T2T Protocol Implementation Details}
\label{sec:protocol-implementation}

The footprint of our protocol implementation was 8\,kB (program memory) and 606\,B (RAM). The complete protocol was implemented in C language, amassing to 2113 lines of code.

\section{Backscatter T2T Network: Evaluation}
\label{sec:results}

We now proceed with describing the experimental results of our backscatter T2T network. We start by introducing the T2T network setup and measurement methodology.

\subsection{Backscatter T2T Network: Experiment Setup}
\label{sec:experiment_setup}

We performed all experiments indoors, in a large office space with many metallic shelves and concrete walls of mixed thickness. A total of seven tags were used in the experiments. Tags and the exciter's antenna were mounted on tripods, elevated 1\,m above ground.

\subsection{\textbf{Result 1}---T2T Range and Multi-hop Performance}
\label{sec:mh-range-extension}

\begin{figure}[t]
	\subfloat[T2T multi-hop range at hop H]{\includegraphics[width=0.5\columnwidth]{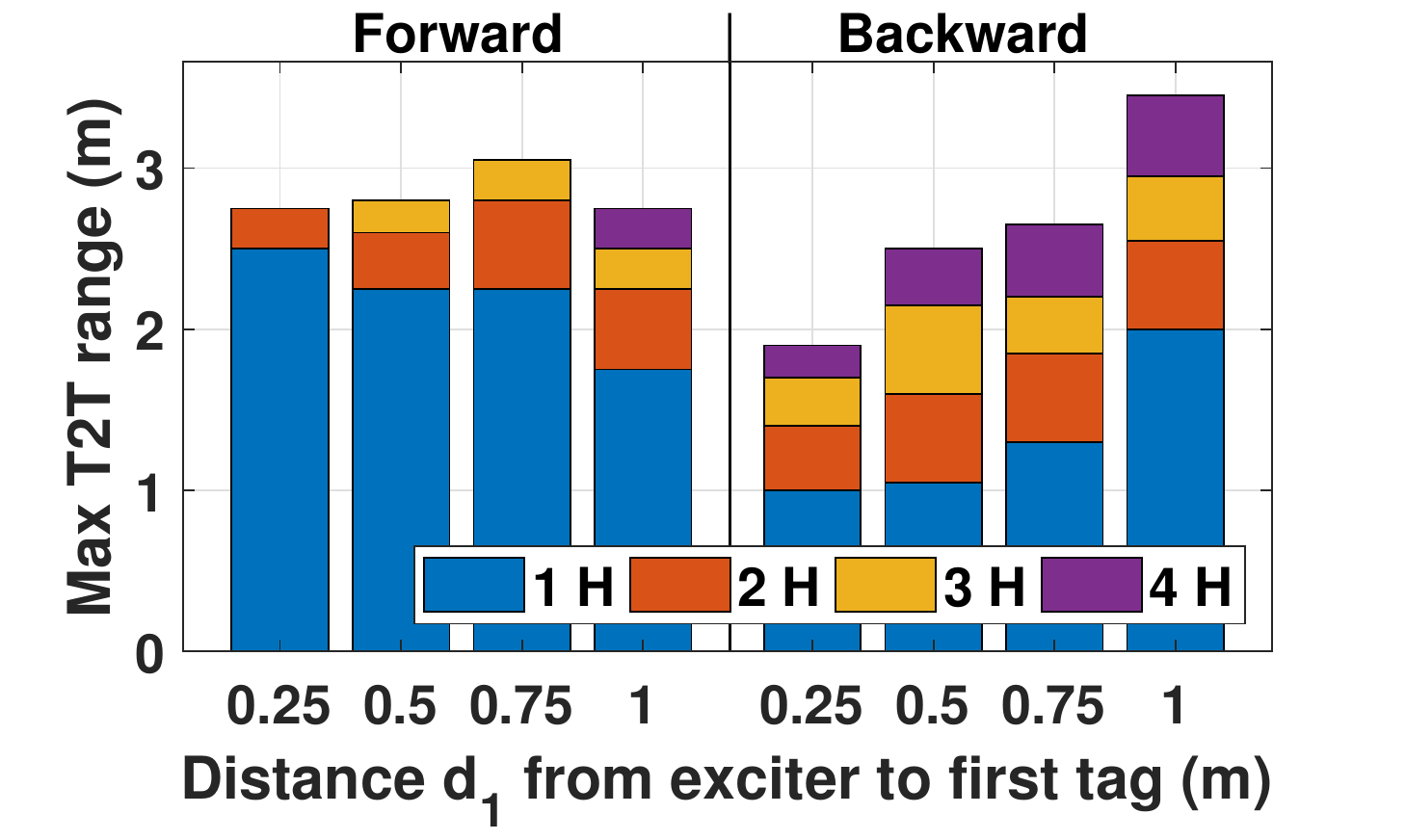}%
		\label{fig:range_extension}}
	\subfloat[Received frame distribution]{\includegraphics[width=0.5\columnwidth]{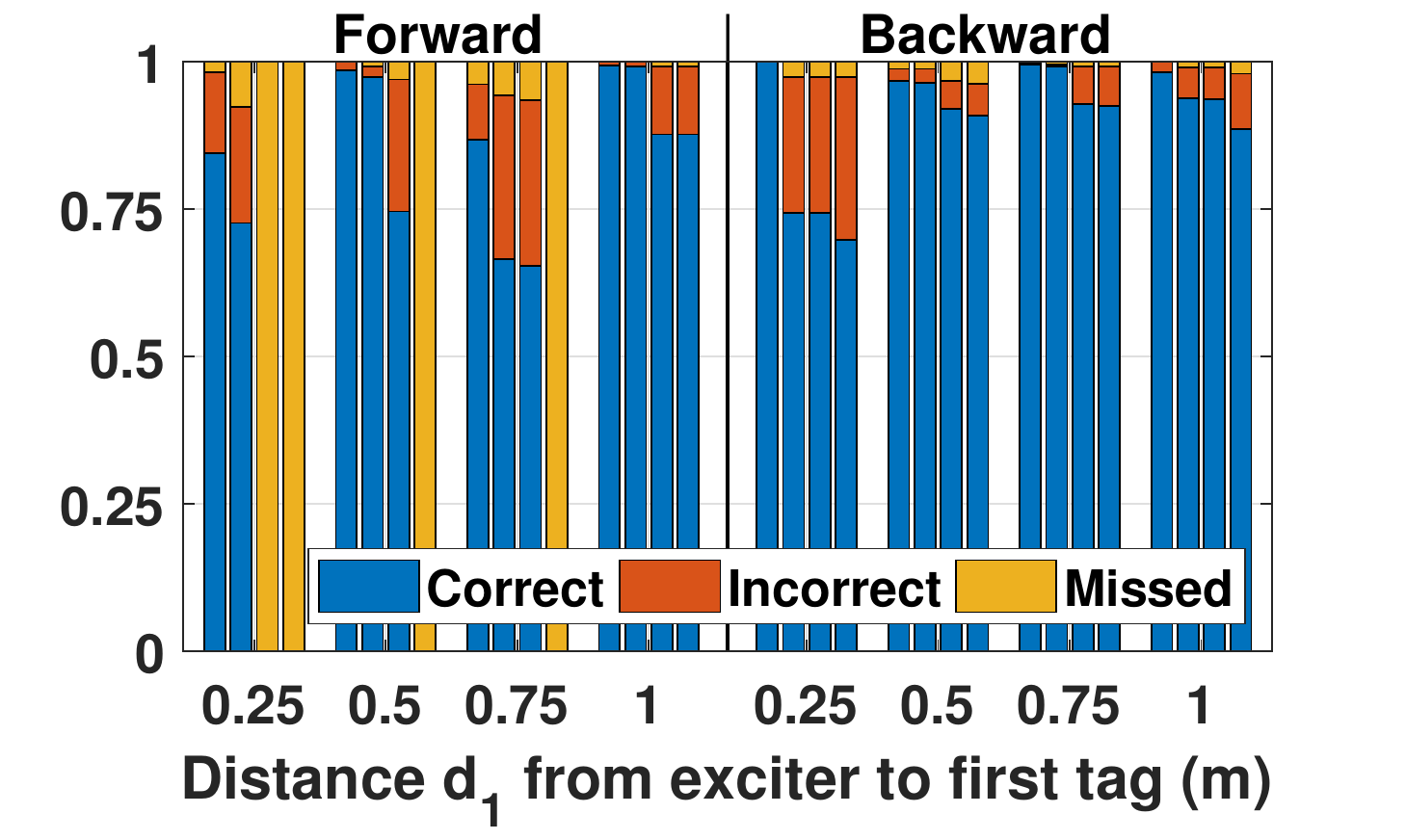}%
		\label{fig:PER_vs_d1}}
	\caption{T2T link metrics for forward and backward link. Fig.~\ref{fig:range_extension}: the backward link has a higher multi-hop gain, \emph{doubling the range} as receiving tag moves away from the exciter; Fig.~\ref{fig:PER_vs_d1}: frame distribution per hop (ordered from left to right in each bar group)---backward link is less prone to errors .}
	\label{fig:range_and_PER}
\end{figure}

\emph{Measurement Methodology.} Each measurement (data point in Fig.~\ref{fig:range_and_PER}) is the average of five runs, each of them consisting of a hundred frames sent by the source tag, see Table~\ref{tab:frame_structure}. The maximum distance of a hop is assumed to be reached when 75\% of the frames are correctly received at the destination.

Fig.~\ref{fig:range_and_PER} presents the core results on the multi-hop T2T link properties. Measurements of the range improvement that multi-hop brings in backscatter T2T networks are given in Fig.~\ref{fig:range_extension}, both for forward and backward links. The figure shows the maximum distance that can be covered in a one-dimensional (line) multi-hop backscatter T2T network as a function of the number of hops and the distance $d_1$ from the exciter to the first T2T tag. The maximum number of hops is shown, meaning that, for instance for $d_1=0.25$\,m only two hops were achievable in the forward link.

The core observation is that the ratio of first versus subsequent hops distance (i.e. multi-hop gain) increases with $d_1$ mainly because the first hop becomes shorter as less power to backscatter is available at the first tag. Furthermore, when T2T tags move away from the exciter, equivalent changes in the distance have less impact on available carrier power due to the logarithmic nature of path loss, increasing both the number and the length of the subsequent hops. As predicted by Observation~\ref{obs:MH_gain}, the multi-hop gain is higher in a backward link, increasing the range by about a factor of two already at four hops.

In Fig.~\ref{fig:PER_vs_d1} the cumulative received frame distribution per hop is shown as a function of $d_1$, for both forward and backward links. Bars are grouped in order of hops, such that the leftmost bar of a group refers to the first hop and the rightmost to the last possible hop for that distance. The main observation here is that the \emph{backward link has a more stable} behavior.


\begin{figure}
	\centering
	\includegraphics[width=\columnwidth]{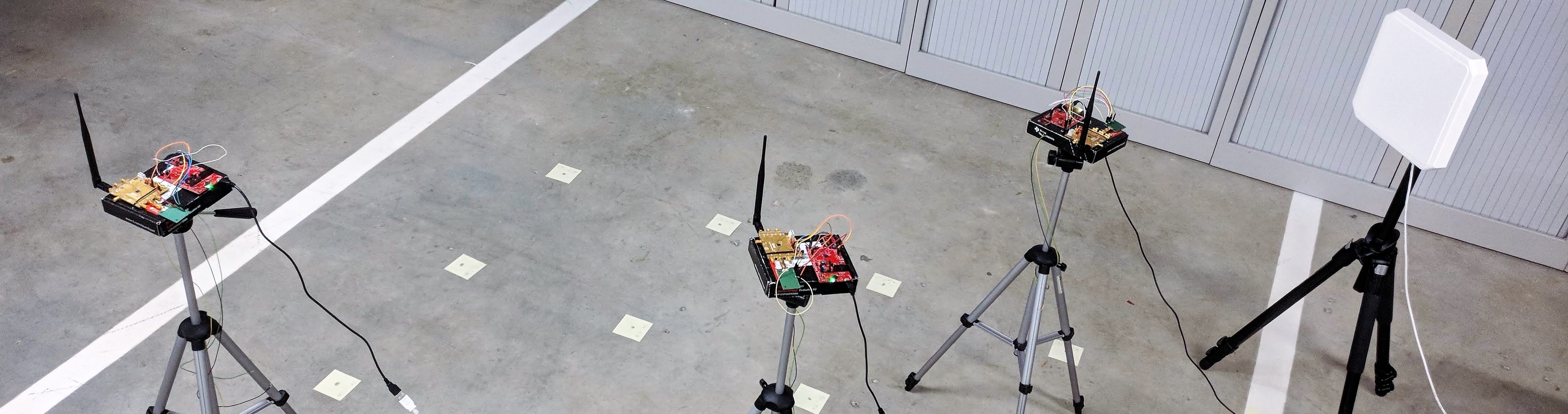}
	\caption{Photograph of the setup to evaluate phase cancellation countermeasures in T2T networks, discussed in Section~\ref{sec:results-phase_cancel_and_forwarding}: Three backscatter T2T tags on tripods, next to white panel antenna; grid coordinates marked with yellow squares on the floor.}
	\label{fig:photo_setup}
\end{figure}

\begin{figure}[t]
	\subfloat[Forward link multi-hop]{\includegraphics[width=0.5\columnwidth]{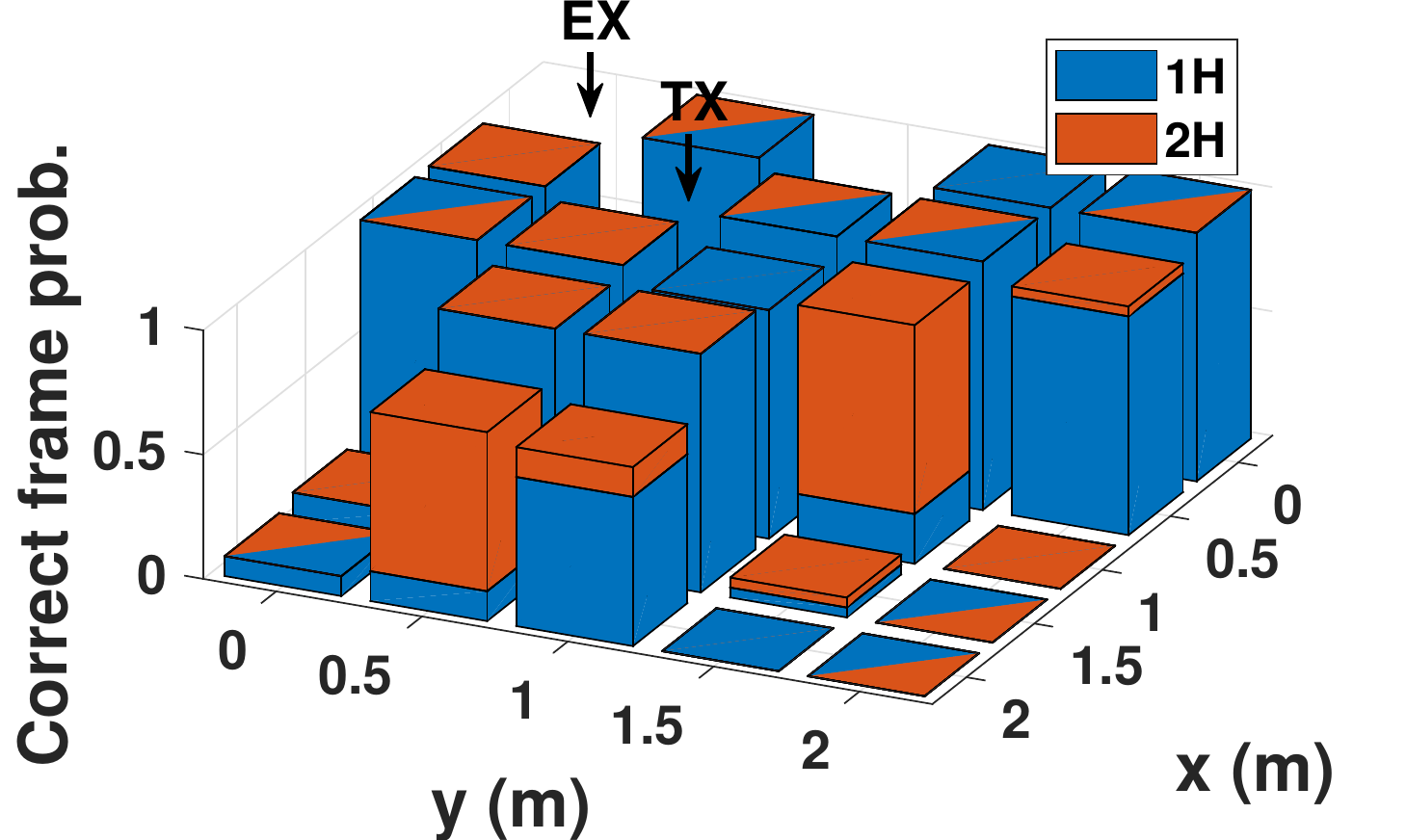}%
		\label{fig:area_MH_FW}}
	\subfloat[Forward link phase shifting]{\includegraphics[width=0.5\columnwidth]{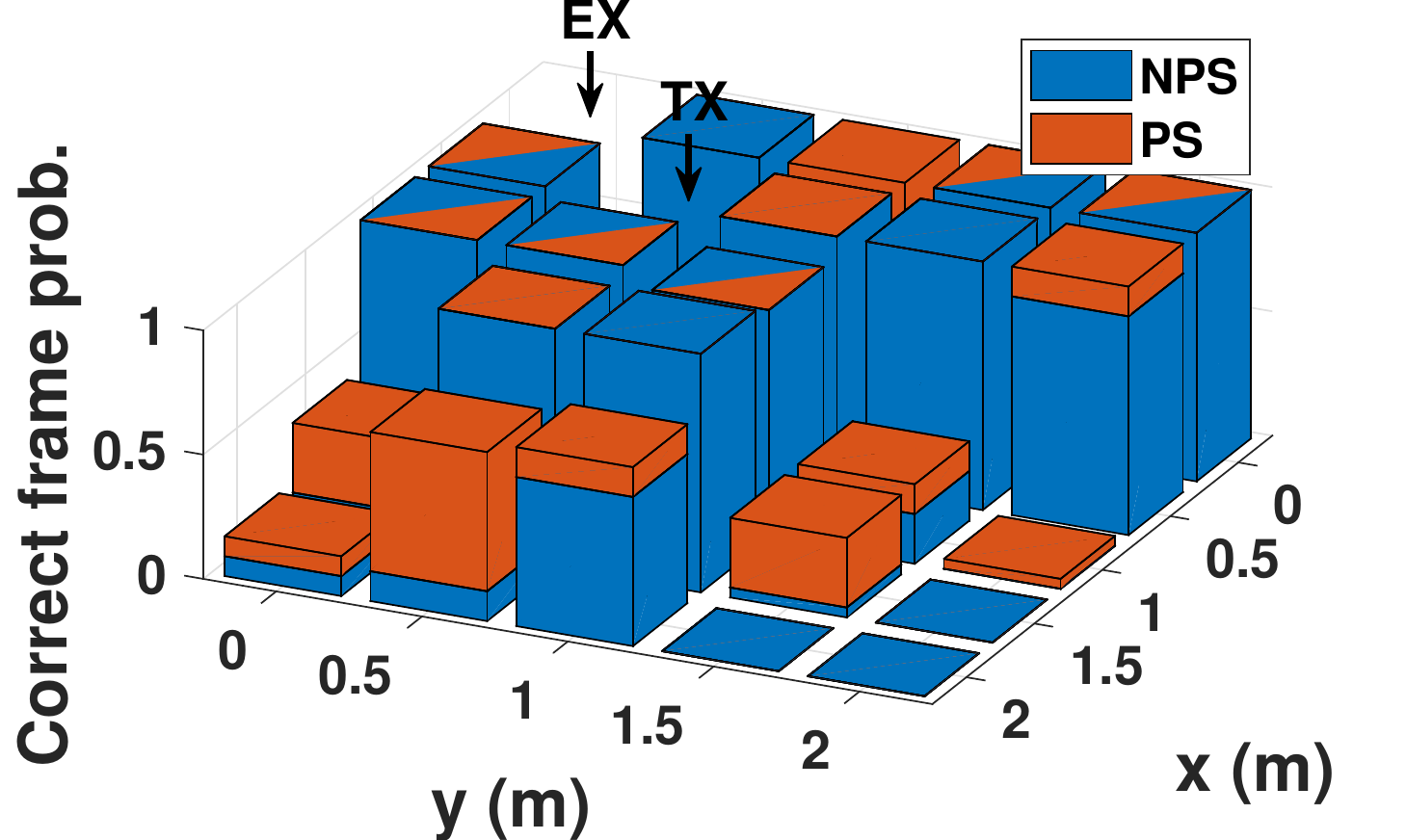}%
		\label{fig:area_PS_FW}}\\	
	\subfloat[Backward link multi-hop]{\includegraphics[width=0.5\columnwidth]{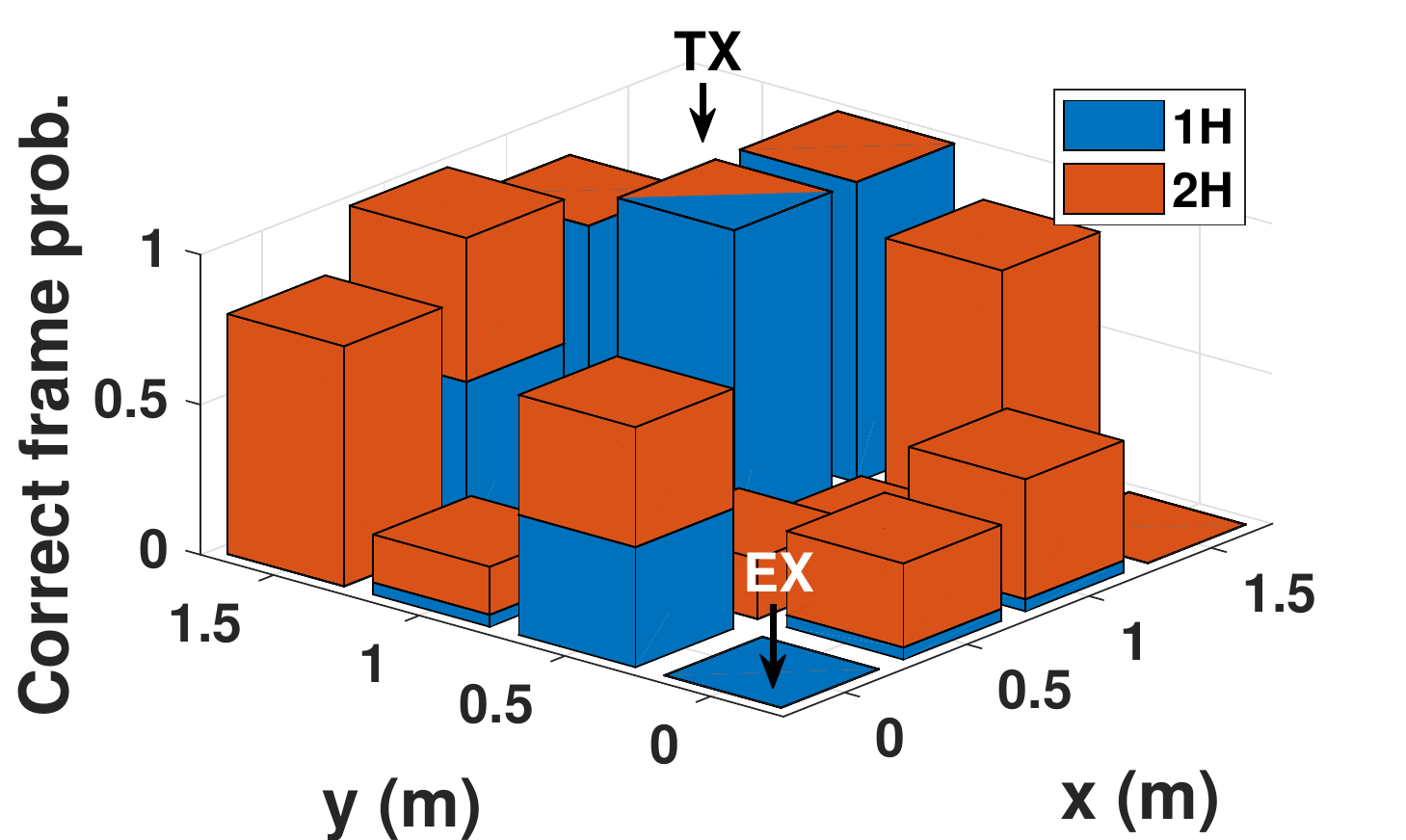}%
		\label{fig:area_MH_BW}}
	\subfloat[Backward link phase shifting]{\includegraphics[width=0.5\columnwidth]{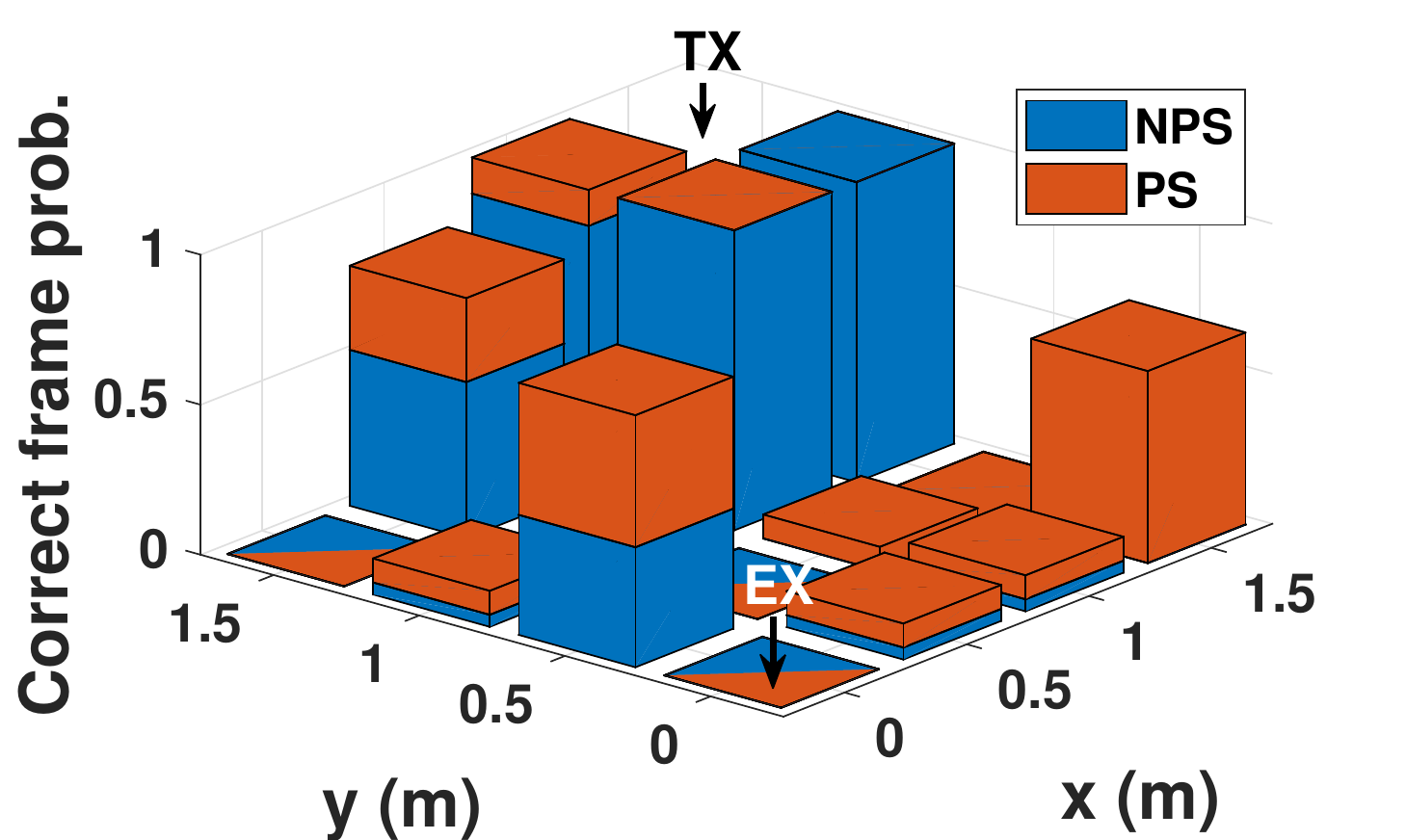}%
		\label{fig:area_PS_BW}}
	\caption{Backscatter T2T network phase cancellation experiment. \emph{EX} and \emph{TX} mark the positions of the exciter and transmitter, respectively. \emph{(N)PS}: (no) phase shifting. Fig.~\ref{fig:area_MH_FW} and~\ref{fig:area_PS_FW}: forward link. Both methods yield comparable results. Fig.~\ref{fig:area_MH_BW} and~\ref{fig:area_PS_BW}: backward link. Multi-hop is superior in network coverage. Note: \emph{1H} and \emph{NPS} is the same case.}
	\label{fig:area_MH_vs_PS}
\end{figure}

\subsection{\textbf{Result 2}---Phase Cancellation and Network Robustness}
\label{sec:results-phase_cancel_and_forwarding}

\emph{Measurement Methodology.} A 2\,m$\times$2\,m area is divided into a grid of 0.5\,m increments. The exciter and source tag are positioned at relative location (0,0)\,m and (0.5,0.5)\,m, respectively. Each data point is the average of three runs of twenty-five frames each.

We now cover the performance of multi-hop as a solution against phase cancellation and compare it to phase shifting for the same purpose (i.e. sending every frame twice with a phase offset of 90$^{\circ}$). This experiment also shows the robustness increase in the T2T network by the use of multi-hop. Furthermore, adding or subtracting relaying tags simulates the case of message forwarding during interference and network reshaping, e.g. from T2T tag mobility.

We proceed by placing a destination tag and measuring the rate of correct frame reception at each coordinate of the grid. Then, we compare this benchmark with the two approaches. The relaying  tags are added at the closest grid coordinates to the middle point between source and destination tags, forming a two-dimensional network, see Fig.~\ref{fig:photo_setup}. For the sake of comparison fairness, this approach only uses one relaying tag (e.g. two hops) so that both solutions are balanced in network utilization: twice the original number of frames in both cases.

\begin{table}
	\centering
	\scriptsize
	\begin{threeparttable}
	\caption{T2T Network Coverage: Summary of Fig.~\ref{fig:area_MH_vs_PS} results}
	\label{table:area_MH_vs_PS}
	\begin{tabular}{r|c|c|c|}
		\cline{2-4}
		& Vanilla\tnote{$\star$}\,\,\,(\%) & Phase shifting\tnote{$\dagger$}\,\,\,(\%) & Multi-hop flooding\tnote{$\dagger$} (\%) \\ \hline
		\multicolumn{1}{|r|}{Forward link}  & 58.1         & 67.3    & 65.0    \\ \hline
		\multicolumn{1}{|r|}{Backward link} & 28.0         & 40.9    & 54.3    \\ \hline
	\end{tabular}

	\begin{tablenotes}
		\item[$\star$] \scriptsize{No phase cancellation fighting mechanism, i.e. traditional single-hop T2T}
		\item[$\dagger$] \scriptsize{As analyzed in Section~\ref{sec:analysis_efficiency}}
	\end{tablenotes}
	\end{threeparttable}
\end{table}

Fig.~\ref{fig:area_MH_vs_PS} presents the results of the experiment, while global (average) coverage values are shown in Table~\ref{table:area_MH_vs_PS}. Note that for the backward link the network area was reduced to 1.5\,m$\times$1.5\,m due to the weaker nature of this link. Points of weak reception in the area could be caused by phase cancellation or by multi-path fading due to the unfavorable testing environment. This hypothesis gains relevance when we look at the cases in which reception is greatly improved by adding a relay tag or by using phase shifting. However, other points could not be enhanced by either method.

In the forward link, multi-hop and phase shifting as means to fight phase cancellation report close results, improving coverage moderately by about 1.13$\times$. On the other hand, the backward link is better handled by the multi-hop solution, almost doubling T2T network coverage. Multi-hop is therefore preferred in either case, as it also adds range extension and robustness to the T2T network.

\subsection{Case Study: Joining Two T2T Backscatter Clusters}
\label{sec:results-case_study}

As a final experiment, we present a case study showcasing the junction of two tag clusters. This study displays the possibility of joining distinct T2T networks with their own exciters by positioning a middle tag connecting both groups.

\begin{figure}
	\centering
	\includegraphics[width=\columnwidth]{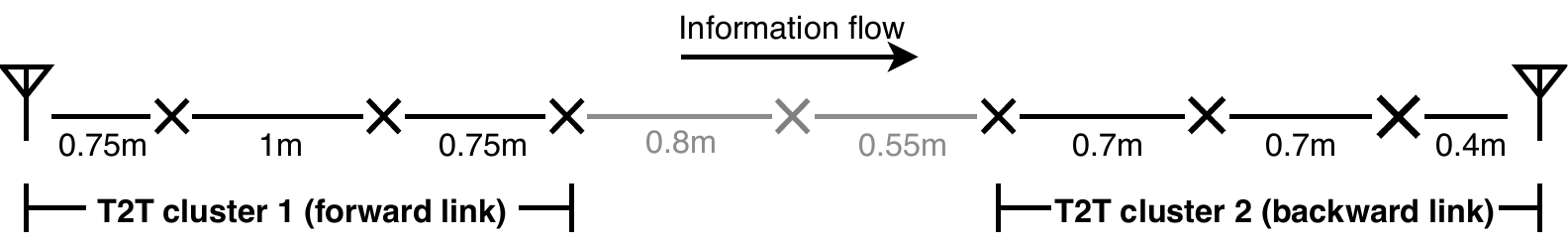}
	\caption{A scenario of T2T clusters junction by adding a bridging T2T node. Antenna symbols represent exciters and crosses represent tags.}
	\label{fig:case_study_cluster}
\end{figure}

The first cluster consists of three tags, spanning a total distance of 2.5\,m from their exciter. At the other side of the laboratory, there is a second cluster, also composed of three tags, reaching 1.8\,m in the other direction. Fig.~\ref{fig:case_study_cluster} depicts this scenario. By placing the seventh tag in the middle of the two groups, we are able to join them and successfully establish communication from one side to the other. There is a forward link in the first cluster, followed by a backward one in the second cluster of tags, covering a total distance of 5.65\,m.

\section{Limitations of this Work}
\label{sec:limitations}


\emph{Hardware Improvements.} Beyond the improvement to increase the receiver sensitivity of the T2T tag, which would increase the T2T communication distance, other developments are required. These include building a full-duplex or multi-antenna transceiver (enabling space-time coding, for instance). 

\emph{Protocol Improvements.} T2T network was not designed with security in mind (to speed up the design time), i.e. transmitted data is not obfuscated. Other improvements include a better (non-flooding) routing algorithm and link adaptation mechanism (based on link packet error rate measurements, for instance). Moreover, we have not performed experiments on harvested power only, so the consideration of T2T networking under intermittent power~\cite{lucia_snapl_2017, Yildirim} is another critical research step.

\section{Conclusions}
\label{sec:conclusions}

In this paper, we built, characterized and evaluated a network composed of backscatter tag-to-tag (T2T) links. In short, we report the first successful demonstration of the largest multi-hop decode-and-relay T2T network using distributed embedded backscatter transceivers. Our results show that T2T multi-hop approach is superior to phase-shifted frame repetition technique in mitigating the effect of dead spots in T2T backscatter networks.   

\section*{Acknowledgments}
We thank NEDAP N.V. (in particular Ben van Zon and Maarten Mijwaart)
and twtg.io (in particular Clement Arnardi and Coen Roest) for
technical support during the hardware design of both versions of
backscatter tag.



\bibliographystyle{IEEEtran}
\bibliography{tag_to_tag_network}

\begin{thebibliography}{10}
\providecommand{\url}[1]{#1}
\csname url@samestyle\endcsname
\providecommand{\newblock}{\relax}
\providecommand{\bibinfo}[2]{#2}
\providecommand{\BIBentrySTDinterwordspacing}{\spaceskip=0pt\relax}
\providecommand{\BIBentryALTinterwordstretchfactor}{4}
\providecommand{\BIBentryALTinterwordspacing}{\spaceskip=\fontdimen2\font plus
\BIBentryALTinterwordstretchfactor\fontdimen3\font minus
  \fontdimen4\font\relax}
\providecommand{\BIBforeignlanguage}[2]{{%
\expandafter\ifx\csname l@#1\endcsname\relax
\typeout{** WARNING: IEEEtran.bst: No hyphenation pattern has been}%
\typeout{** loaded for the language `#1'. Using the pattern for}%
\typeout{** the default language instead.}%
\else
\language=\csname l@#1\endcsname
\fi
#2}}
\providecommand{\BIBdecl}{\relax}
\BIBdecl

\bibitem{talla_arxiv_2017}
V.~{Talla} \emph{et~al.}, ``{LoRa} backscatter: Enabling the vision of
  ubiquitous connectivity,'' \emph{Proc. ACM Interact. Mob. Wearable Ubiquitous
  Technol.}

\bibitem{zhang_sigcomm_2016}
P.~{Zhang} \emph{et~al.}, ``Enabling practical backscatter for on-body
  sensors,'' in \emph{Proc. ACM SIGCOMM}, 2016.

\bibitem{alevizos_arxiv_2017}
\BIBentryALTinterwordspacing
P.~N. {Alevizos} \emph{et~al.}, ``Multistatic scatter radio sensor networks for
  extended coverage,'' \emph{{IEEE} Trans. Wireless Commun.}, 2018. [Online].
  Available: \url{https://arxiv.org/abs/1706.03091}
\BIBentrySTDinterwordspacing

\bibitem{griffin_apm_2009}
J.~D. {Griffin} and G.~D. {Durgin}, ``Complete link budgets for
  backscatter-radio and {RFID} systems,'' \emph{{IEEE} Antennas Propagat.
  Mag.}, 2009.

\bibitem{karimi_rfid_2017}
Y.~{Karimi} \emph{et~al.}, ``Design of a backscatter-based tag-to-tag system,''
  in \emph{Proc. IEEE RFID}, 2017.

\bibitem{nikitin_rfid_2012}
P.~V. {Nikitin} \emph{et~al.}, ``Passive tag-to-tag communication,'' in
  \emph{Proc. IEEE RFID}, 2012.

\bibitem{parks_sigcomm_2014}
A.~N. {Parks} \emph{et~al.}, ``Turbocharging ambient backscatter
  communication,'' in \emph{Proc. ACM SIGCOMM}, 2014.

\bibitem{liu_jsac_2015}
Q.~{Liu} \emph{et~al.}, ``Green wireless power transfer networks,''
  \emph{{IEEE} J. Select. Areas Commun.}, 2016.

\bibitem{ti_msp430_website}
\BIBentryALTinterwordspacing
{Texas Instruments}. (2017) {MSP430FR5969} 16\,{MHz} ultra-low-power
  microcontroller product page. [Online]. Available:
  \url{http://www.ti.com/product/MSP430FR5969}
\BIBentrySTDinterwordspacing

\bibitem{tag-to-tag_source_files}
\BIBentryALTinterwordspacing
(2019) Hardware/software source files of {T2T} backscatter network. [Online].
  Available: \url{https://github.com/TUDSSL/Backscatter-Network}
\BIBentrySTDinterwordspacing

\bibitem{gummeson_mobisys_2012}
J.~{Gummeson} \emph{et~al.}, ``Flit: A bulk transmission protocol for
  {RFID}-scale sensors,'' in \emph{Proc. ACM MobiSys}, 2012.

\bibitem{wang_sigcomm_2012}
J.~{Wang} \emph{et~al.}, ``Efficient and reliable low-power backscatter
  networks,'' in \emph{Proc. ACM SIGCOMM}, 2012.

\bibitem{zhang_mobisys_2012}
P.~{Zhang} \emph{et~al.}, ``{BLINK}: A high throughput link layer for
  backscatter communication,'' in \emph{Proc. ACM Mobisys}, 2012.

\bibitem{yang_infocom_2017}
C.~{Yang} \emph{et~al.}, ``Riding the airways: Ultra­wideband ambient
  backscatter via commercial broadcast systems,'' in \emph{Proc. IEEE INFOCOM},
  2017.

\bibitem{zhang_mobicom_2014}
P.~{Zhang} \emph{et~al.}, ``Ekhonet: High speed ultra low-power backscatter for
  next generation sensors,'' in \emph{Proc. ACM MobiCom}, 2014.

\bibitem{liu_sigcomm_2013}
V.~{Liu} \emph{et~al.}, ``Ambient backscatter: Wireless communication out of
  thin air,'' in \emph{Proc. ACM SIGCOMM}, 2013.

\bibitem{shen_iotj_2016}
Z.~{Shen} \emph{et~al.}, ``Phase cancellation in backscatter-based tag-to-tag
  communication systems,'' \emph{{IEEE} Internet Things J.}, 2016.

\bibitem{nikitin_rfid_2011}
P.~V. {Nikitin} \emph{et~al.}, ``{RFID} paperclip tags,'' in \emph{Proc. IEEE
  RFID}, 2011.

\bibitem{barnet_mobisys}
J.~{Ryoo} \emph{et~al.}, ``{BARNET}: Towards activity recognition using passive
  backscattering tag-to-tag network,'' in \emph{Proc. ACM MobiSys}, 2018.

\bibitem{bttn_iot}
J.~Ryoo \emph{et~al.}, ``Design and evaluation of ‘bttn’: A backscattering
  tag-to-tag network,'' \emph{IEEE Internet of Things Journal}, 2018.

\bibitem{liu_awpl_2011}
H.-C. {Liu} \emph{et~al.}, ``Passive {UHF} {RFID} tag with backscatter
  diversity,'' \emph{{IEEE} Antennas Wireless Propagat. Lett.}, 2011.

\bibitem{ads_website}
\BIBentryALTinterwordspacing
{Keysight}. (2018) {Advanced Design System} website. [Online]. Available:
  \url{www.keysight.com/find/eesof-ads}
\BIBentrySTDinterwordspacing

\bibitem{orcad_website}
\BIBentryALTinterwordspacing
{Cadence Design Systems}. (2018) {OrCAD Capture} application website. [Online].
  Available: \url{http://www.orcad.com/products/orcad-capture}
\BIBentrySTDinterwordspacing

\bibitem{hp_note_963}
\BIBentryALTinterwordspacing
{Hewlett-Packard Co.} (1980) Impedance matching techniques for mixers and
  detectors (application note 963). [Online]. Available:
  \url{http://67.225.133.110/~gbpprorg/mil/inter/an963.pdf}
\BIBentrySTDinterwordspacing

\bibitem{SLMD121H04L_website}
\BIBentryALTinterwordspacing
{IXYS Corporation}. (2018) {IXOLAR}\texttrademark~high efficiency {SLMD121H04L}
  solar module. [Online]. Available:
  \url{http://ixapps.ixys.com/DataSheet/SLMD121H04L_Nov16.pdf}
\BIBentrySTDinterwordspacing

\bibitem{BQ25570EVM-206_website}
\BIBentryALTinterwordspacing
{Texas Instruments}. (2018) Ultra low power management {IC}, boost charger
  nanopowered buck converter evaluation module. [Online]. Available:
  \url{http://www.ti.com/tool/BQ25570EVM-206}
\BIBentrySTDinterwordspacing

\bibitem{BQ25570_website}
\BIBentryALTinterwordspacing
T.~Instruments. (2018) Ultra low power harvester power management {IC} with
  boost charger, and nanopower buck converter. [Online]. Available:
  \url{http://www.ti.com/product/BQ25570}
\BIBentrySTDinterwordspacing

\bibitem{EXP430FR5969_website}
\BIBentryALTinterwordspacing
{Texas Instruments}. (2018) {MSP430FR5969} launchpad development kit. [Online].
  Available: \url{http://www.ti.com/tool/MSP-EXP430FR5969}
\BIBentrySTDinterwordspacing

\bibitem{etsi_standard_302_208}
\BIBentryALTinterwordspacing
{ETSI}, ``Radio frequency identification equipment operating in the band
  865\,{MHz} to 868\,{MHz} with power levels up to 2\,{W} and in the band
  915\,{MHz} to 921\,{MHz} with power levels up to 4\,{W},'' 2016. [Online].
  Available:
  \url{https://www.etsi.org/deliver/etsi_en/302200_302299/302208/03.01.01_60/en_302208v030101p.pdf}
\BIBentrySTDinterwordspacing

\bibitem{monsoon}
\BIBentryALTinterwordspacing
{Monsoon Solutions Inc.}, ``High voltage power monitor,'' 2018. [Online].
  Available: \url{https://www.msoon.com/hvpm-product-documentation}
\BIBentrySTDinterwordspacing

\bibitem{ublox_saran2}
\BIBentryALTinterwordspacing
u~blox. (2018) Sara-n2 power-optimized nb-iot (lte cat nb1) modules data sheet.
  [Online]. Available:
  \url{https://www.u-blox.com/sites/default/files/SARA-N2_DataSheet_%28UBX-15025564%29.pdf}
\BIBentrySTDinterwordspacing

\bibitem{chen2013range}
L.~{Chen} \emph{et~al.}, ``Range extension of passive wake-up radio systems
  through energy harvesting,'' in \emph{Proc. IEEE ICC}, 2013.

\bibitem{aantjes_hlpc_2016}
\BIBentryALTinterwordspacing
K.~S. Y{\i}ld{\i}r{\i}m \emph{et~al.} (2016) On the synchronization of
  intermittently powered wireless embedded systems. [Online]. Available:
  \url{https://arxiv.org/abs/1606.01719}
\BIBentrySTDinterwordspacing

\bibitem{polastre_sensys_2004}
J.~{Polastre} \emph{et~al.}, ``Versatile low power media access for wireless
  sensor networks,'' in \emph{Proc. ACM SenSys}, 2004.

\bibitem{lucia_snapl_2017}
B.~{Lucia} \emph{et~al.}, ``Intermittent computing: Challenges and
  opportunities,'' in \emph{Proc. SNAPL}, 2017.

\bibitem{Yildirim}
K.~S. Y{\i}ld{\i}r{\i}m \emph{et~al.}, ``Ink: Reactive kernel for tiny
  batteryless sensors,'' in \emph{In Proc. of SenSys}.\hskip 1em plus 0.5em
  minus 0.4em\relax ACM, 2018, pp. 41--53.

\end{thebibliography}

\end{document}